\documentclass[11pt]{article}
\usepackage{amsfonts,amsmath,amssymb}
\usepackage{enumerate}
\usepackage[hidelinks]{hyperref}
\usepackage{bbm}
\usepackage{nicefrac}
\usepackage[all]{xy}
\usepackage{graphicx}
\usepackage{bm}
\usepackage{makecell}
\usepackage{tabularx}
\usepackage{caption}
\usepackage[table]{xcolor}
\usepackage[titles]{tocloft}
\usepackage{upgreek}
\usepackage{cancel}			
\usepackage{multicol}		
\usepackage{float}			
\usepackage{fontawesome}	
\usepackage{arydshln}		

\usepackage{framed}

\usepackage{longtable}      

\usepackage{booktabs}

\renewcommand{\baselinestretch}{1.1}

\newcommand{\be}{\begin{equation}}
\newcommand{\ee}{\end{equation}}

\newcommand{\bea}{\begin{eqnarray}}
\newcommand{\eea}{\end{eqnarray}}

\newcommand{\bes}{\begin{subequations}}
\newcommand{\ees}{\end{subequations}}

\newcommand{\cN}{{\cal N}}

\def\sst#1{{\scriptscriptstyle #1}}

\def\0{{\sst{(0)}}}
\def\1{{\sst{(1)}}}
\def\2{{\sst{(2)}}}
\def\3{{\sst{(3)}}}
\def\4{{\sst{(4)}}}
\def\5{{\sst{(5)}}}
\def\6{{\sst{(6)}}}
\def\7{{\sst{(7)}}}
\def\8{{\sst{(8)}}}

\def\cV{{{\cal V}}}
\def\cM{{{\cal M}}}

\newcommand{\cL}{{\cal L}}

\newcommand{\vol}{\textrm{vol}}

\allowdisplaybreaks  

\usepackage{multirow}
\usepackage{rotating}

\def\d{\textrm{d}}

\newcommand{\ba}{\begin{align}}
\newcommand{\ea}{\end{align}}

\newcommand{\bse}{\begin{subequations}}
\newcommand{\ese}{\end{subequations}}

\allowdisplaybreaks

\newcommand{\dd}{\mathrm d}
\newcommand{\T}{\mathsf T}
\newcommand{\diag}{\operatorname{diag}}
\newcommand{\HH}{\mathcal H}
\newcommand{\one}{\mathbf 1}

\begin{document}

\makeatletter
\renewcommand{\theequation}{\thesection.\arabic{equation}}
\@addtoreset{equation}{section}
\makeatother

\begin{titlepage}

\begin{flushright}


\end{flushright}

\vspace{5pt}

   \begin{center}
   \baselineskip=16pt

\begin{Large}\textbf{
All known supersymmetric $\mathrm{AdS}_4 \times S^7$ uplifts \\[10pt] 
 from maximal supergravity} 
   \end{Large}

\vspace{25pt}

{\large Krzysztof Pilch$^{1}$ \,and \,  Oscar Varela$^{2,3}$}
		
\vspace{45pt}

	\begin{small}

	{\it $^{1}$ Department of Physics and Astronomy, University of Southern California, \\ Los Angeles, CA 90089, USA}     \\	

	\vspace{15pt}
	
	{\it $^{2}$ Quantum Gravity and Interactions Initiative, \\ Department of Physics, Utah State University, Logan, UT 84322, USA}     \\	
	
	\vspace{15pt}

	{\it $^{3}$ Instituto de F\'\i sica Te\'orica UAM/CSIC, 28049 Madrid, Spain}  \\

	\end{small}

\vskip 50pt

\end{center}

\begin{center}
\textbf{Abstract}
\end{center}

\begin{quote}

The solutions of $D=4$ $\mathcal N=8$ SO(8)-gauged supergravity uplift consistently to eleven dimensions. All known supersymmetric AdS$_4$ vacua of this theory lie in an $\mathcal N=1$ truncation containing seven chiral multiplets. In this paper, we work out the uplift of this seven-chiral sector, providing a unified eleven-dimensional geometric description of all known supersymmetric AdS$_4$ solutions in this class. We then specialise to the $\mathcal N=1$ vacua with U(1)$^2$ and SO(3) symmetry, obtain their new $D=11$ uplifts and bring out their internal geometries. Both of them  admit a Kaluza--Klein-type description. For the U(1)$^2$ solution, this uses the global Hopf fibration of $S^7$ over $\mathbb{CP}^3$. One U(1) acts on the 
fibre, while the other generates the only continuous isometry of the
deformed $\mathbb{CP}^3$ base. For the SO(3) solution, the geometry takes
a local non-abelian Kaluza--Klein form, with SO(3) acting on
the fibres and no residual continuous symmetry on the
four-dimensional base. 

\end{quote}

\vfill

\end{titlepage}

\tableofcontents



\section{Introduction}
\label{sec:introduction}

All solutions of maximal $D=4$ SO(8)-gauged supergravity \cite{deWit:1981sst,deWit:1982ig} can be uplifted to eleven-dimensional supergravity \cite{ Cremmer:1978km} through the consistent reduction on $S^7$ \cite{deWit:1983vq,deWit:1984nz,deWit:1986iy,Duff:1986hr,Nicolai:2011cy,deWit:2013ija,Godazgar:2013nma,Lee:2014mla,Varela:2015ywx,Kruger:2016agp}. In particular, the $D=4$ AdS vacua (see \cite{Comsa:2019rcz} for a survey) give rise to AdS$_4\times S^7$ backgrounds, with a generally warped and deformed  geometry on $S^7$. Five supersymmetric $D=4$ vacua are known (see table~\ref{tab:AllSusyAdS4} for a summary): the maximally supersymmetric SO(8) point \cite{deWit:1982ig}, the $\mathcal N=2$ SU(3)$\times$U(1) point \cite{Warner:1983vz}, and three $\mathcal N=1$ points with G$_2$ \cite{Warner:1983vz}, U(1)$^2$ \cite{Fischbacher:2009cj,Fischbacher:2010ec} and SO(3) \cite{Comsa:2019rcz,Bobev:2019dik} symmetry. Explicit eleven-dimensional descriptions have long been available for the first two and the G$_2$ point: these are the round Freund--Rubin background \cite{Freund:1980xh}, the SU(3)$\times$U(1)-invariant (CPW) solution of \cite{Corrado:2001nv}, and the G$_2$-invariant solution of \cite{deWit:1984nz,Godazgar:2013nma}. Here we close the remaining gap by providing the uplifts of the U(1)$^2$ and SO(3) vacua and bringing out their internal geometries. Our starting point is the observation that all five vacua occur within the same $\mathcal N=1$ sector containing seven chiral multiplets \cite{Bobev:2019dik}. By uplifting this sector as a whole, we obtain an overarching eleven-dimensional description of all known supersymmetric vacua, which we then specialise to the two remaining solutions.

The $D=4$ seven-chiral model has an appealing connection with the Fano plane \cite{Bobev:2019dik}. Recall that this is the finite geometry with seven points and seven lines, with three points on each line. The seven complex scalars can be assigned to its points, and the cubic terms in the superpotential to its lines. We will find that the same organisation extends to the uplift: the lines and their complements arrange the scalar dependence of the internal metric and three-form potential. This provides a useful way of keeping otherwise lengthy expressions under control. There is, however, a geometric aspect that deserves separate attention. The model is obtained by retaining the fields invariant under a discrete $\mathbb Z_2^3$ subgroup of SO(8) \cite{Bobev:2019dik}. Such a truncation does not in general make the continuous symmetries of its individual vacua manifest, and its uniform scalar description can obscure the rather different local internal geometries. We will therefore supplement the common uplift formulae with descriptions adapted to the continuous symmetries of the solutions.

\begin{table}
\centering
\small
\setlength{\tabcolsep}{6pt}
\renewcommand{\arraystretch}{1.5}
\begin{tabular*}{0.92\linewidth}{@{\extracolsep{\fill}}clccc@{}}
\toprule
Supersymmetry & Symmetry $G$ & $g^{-2}V_0$
& \makecell{$D=4$\\solution} & \makecell{$D=11$\\uplift} \\
\midrule
$\mathcal N=8$ & SO(8) & $-24$ & \cite{deWit:1982ig} & \cite{Freund:1980xh} \\
$\mathcal N=2$ & SU(3)$\times$U(1) & $-18\sqrt3$ & \cite{Warner:1983vz} & \cite{Corrado:2001nv} \\
\addlinespace[2pt]
$\mathcal N=1$ & G$_2$ & $-\dfrac{2^{11/2}\,3^{13/4}}{5^{5/2}}$ & \cite{Warner:1983vz} & \cite{deWit:1984nz,Godazgar:2013nma} \\
\addlinespace[2pt]
$\mathcal N=1$ & SO(3) & $-55.363855$ & \cite{Comsa:2019rcz,Bobev:2019dik} & \hyperref[sec:SO3Sol11D]{[here]} \\
$\mathcal N=1$ & U(1)$\times$U(1) & $-48$ & \cite{Fischbacher:2009cj,Fischbacher:2010ec} & \hyperref[sec:U12Sol11D]{[here]} \\
\bottomrule
\end{tabular*}
\caption{\footnotesize{All known supersymmetric AdS$_4$ solutions that consistently uplift to $D=11$ supergravity from maximal $D=4$ SO(8)-gauged supergravity. 
}\normalsize}
\label{tab:AllSusyAdS4}
\end{table}

Recall that the unit seven-sphere is the locus
\begin{equation}\label{eq: S7inR8}
\delta_{IJ}\mu^I\mu^J=1\; ,
\end{equation}
in $\mathbb{R}^8$, with $\mu^I$, $I=1,\dots,8$, transforming in the
$\bm{8}_v$ of SO(8). Its induced round metric,
\begin{equation}\label{eq:RoundS7Met}
\dd s_7^2=\delta_{IJ}\,\dd\mu^I\,\dd\mu^J\; ,
\end{equation}
inherits the SO(8) symmetry of (\ref{eq: S7inR8}). An angular
parametrisation can make only a subgroup of this symmetry manifest.
For example, the choice (\ref{eq:U12CLPcoordinates}) below puts the
round metric (\ref{eq:RoundS7Met}) into the form
\begin{equation}
\dd s_7^2
=\dd s_{\mathrm{FS}}^2(\mathbb{CP}^3)
+(\dd\psi+\mathcal A_{\mathrm{FS}})^2,
\label{eq:introRoundHopf}
\end{equation}
adapted to the Hopf fibration of $S^7$ over $\mathbb{CP}^3$.
Here $\psi$ parametrises the fibre, while
$\dd s_{\mathrm{FS}}^2(\mathbb{CP}^3)$ and $\mathcal A_{\mathrm{FS}}$
are the Fubini--Study metric and Hopf connection. In the conventions
of \cite{Cvetic:2000yp}, these read
\begin{equation}
\begin{aligned}
\dd s_{\mathrm{FS}}^2(\mathbb{CP}^3)={}&\dd\xi^2
+\frac{\cos^2\xi}{4}
 (\dd\theta^2+\sin^2\theta\,\dd\phi^2)
+\frac{\sin^2\xi}{4}
 (\dd\widetilde\theta^2+\sin^2\widetilde\theta\,\dd\widetilde\phi^2)\\
&+\frac{\cos^2\xi\sin^2\xi}{4}
 \bigl(\dd\widetilde\psi-\cos\theta\,\dd\phi
 +\cos\widetilde\theta\,\dd\widetilde\phi\bigr)^2,
\end{aligned}
\label{eq:U12FSmetric}
\end{equation}
and
\begin{equation}
\mathcal A_{\mathrm{FS}}
=\frac14\cos(2\xi)\,\dd\widetilde\psi
-\frac12\cos^2\xi\cos\theta\,\dd\phi
-\frac12\sin^2\xi\cos\widetilde\theta\,\dd\widetilde\phi,
\label{eq:U12Hopfconnection}
\end{equation}
with $\dd\mathcal A_{\mathrm{FS}}=2J_{\mathrm{FS}}$, and
$J_{\mathrm{FS}}$ the K\"ahler form on $\mathbb{CP}^3$.
Although (\ref{eq:introRoundHopf}) retains the full SO(8) symmetry
of (\ref{eq:RoundS7Met}), only its
$\mathrm{SU}(4)\times\mathrm{U}(1)$ subgroup is manifest in this form. Metrics of the form \eqref{eq:introRoundHopf}, in which a U$(1)$ fibre
is coupled to the base through a connection, are commonly said to be
in abelian Kaluza--Klein (KK) form, see \cite{Duff:1986hr}.

The $\cN=8$ SO(8)-invariant vacuum uplifts to the Freund--Rubin
solution \cite{Freund:1980xh}, whose internal metric is proportional
to the round metric (\ref{eq:RoundS7Met}). The other supersymmetric
vacua of $D=4$ $\cN=8$ SO(8)-gauged supergravity uplift to $D=11$
solutions with typically deformed $S^7$ metrics whose isometry groups are
proper subgroups of SO(8). A useful guide is provided by the geometries already encountered in the SU(3)-invariant sector of the $D=4$ supergravity, whose common $D=11$ embedding was developed in \cite{Larios:2019kbw}. The $\mathcal N=1$ G$_2$ solution \cite{deWit:1984nz,Godazgar:2013nma} describes a warped seven-sphere that can be viewed as round six-spheres whose size varies with latitude. Its metric has SO(7) symmetry, while the internal flux, expressed through the nearly-K\"ahler forms on $S^6$, leaves only G$_2$ symmetry \cite{Larios:2019kbw}. The $\mathcal N=2$ CPW solution \cite{Corrado:2001nv} has a different geometric description: its squashed and stretched seven-sphere is naturally written using the Hopf fibration of $S^5$ over $\mathbb{CP}^2$, equipped with the Fubini--Study metric. Here too the flux is essential: the metric admits two commuting circle symmetries, but the full solution preserves only SU(3)$\times$U(1) \cite{Corrado:2001nv}. 

We will present similarly useful descriptions of the uplifts of the remaining two $\mathcal N=1$ vacua, with U(1)$^2$ and SO(3) symmetry. The internal geometries of our two new solutions turn out to admit a KK description, as defined above, adapted to their continuous symmetries. For the U(1)$^2$-invariant solution, the KK form is abelian and uses the same Hopf circle bundle over $\mathbb{CP}^3$ as in \eqref{eq:introRoundHopf}. The metric reads
\begin{equation}
\dd s_7^2
= g^{-2}\Delta^2
  \left[\dd s_6^2+f\,(\dd\psi+\mathcal A)^2\right] ,
\label{eq:introU12KK}
\end{equation}
where $g$ is a constant, $\Delta $ a warp factor, $\dd s_6^2$ and $\mathcal A$ deform the Fubini--Study metric (\ref{eq:U12FSmetric})
and the Hopf connection (\ref{eq:U12Hopfconnection}), and the fibre is squashed by a function $f$. All these quantities are independent of the fibre angle $\psi$ and a base angle $\widetilde\psi$, rendering (\ref{eq:introU12KK}) U$(1)^2$-invariant. For the SO(3)-invariant solution, the internal metric takes the
non-abelian KK form
\begin{equation}
\dd s_7^2
=g^{-2}\Delta^2\left[h_{ab}\,\dd q^a\dd q^b
  +\gamma_{ij}
   \bigl(\sigma^i+\mathcal A^i\bigr)
   \bigl(\sigma^j+\mathcal A^j\bigr)\right],
\qquad
\mathcal A^i=\mathcal A^i{}_a\,\dd q^a .
\label{eq:introSO3KK}
\end{equation}
Here, $q^a$, $a=1, \ldots , 4$, are coordinates on the local four-dimensional base, $\sigma^i$, $i=1,2,3$, are left-invariant one-forms on the SO(3) fibres, and the metric coefficients $h_{ab}$ and $\gamma_{ij}$, as well as the connection coefficients $\mathcal A^i{}_a$, depend only on $q^a$. At the new vacuum, SO(3) acts entirely along the fibres, leaving no residual continuous
symmetry on the base. The abelian description \eqref{eq:introU12KK}
uses the global Hopf fibration of $S^7$, whereas \eqref{eq:introSO3KK} holds only locally. We compute these uplifts using the formulae of
\cite{Varela:2015ywx}, see also \cite{Kruger:2016agp}, specialised
to the seven-chiral sector and then to the U(1)$^2$ and SO(3) vacua. All our results are analytic, and extensive numerical checks confirm that both new configurations satisfy the field equations of $D=11$ supergravity.

The rest of the paper is organised as follows. In section~\ref{sec:7CS4D} we review the seven-chiral sector of maximal
$D=4$ SO(8)-gauged supergravity, its supersymmetric vacua and some 
relevant subsectors. In section~\ref{sec:7CS11D} we
derive the eleven-dimensional warp factor, internal metric and
three-form potential for the full seven-chiral sector. Section~\ref{sec:U12Sol11D} specialises the uplift to a suitable subsector 
with two real scalars. At the scalar origin
we recover the Freund-Rubin solution, 
while the other supersymmetric vacuum yields the new U(1)$^2$ solution.
In section~\ref{sec:SO3Sol11D} we obtain the new SO(3) solution from
the SO(3)-invariant three-chiral subsector, whose uplift also reproduces
the round, G$_2$ and CPW solutions at their respective vacua. We conclude
in section~6 with a discussion of applications and possible extensions.
The appendices collect our conventions, the building blocks of the seven-chiral uplift
and the explicit angular expressions for the internal metrics and
three-form potentials of the two new solutions.


\section{The seven-chiral sector in $D=4$} \label{sec:7CS4D}



Let us first review the seven-chiral subsector \cite{Bobev:2019dik} of $D=4$ $\cN=8$ SO(8)-gauged supergravity, including its supersymmetric vacua and some relevant symmetry enhancements. In the conventions that we are using, summarised in appendix~\ref{app:conventions}, the bosonic sector of the model contains the $D=4$ metric along with seven scalars, $\varphi_i$, $i=1, \ldots , 7$, and seven pseudoscalars, $\chi_i$,  parametrising an $\big( \textrm{SL}(2,\mathbb R)/ \textrm{SO}(2) \big)^7$ submanifold of $\textrm{E}_{7(7)}/\textrm{SU}(8)$. These can be conveniently packed into seven complex fields
\begin{equation} \label{eq:7ComplelsScal}
t_i = -\chi_i + i e^{-\varphi_i} \; , \qquad i= 1 , \ldots , 7 \; ,
\end{equation}
taking values in seven copies of the upper half-plane. For later reference, it is also helpful to introduce the real combinations
\begin{equation} \label{eq:scalars}
a_i=e^{-\varphi_i}+\chi_i^2e^{\varphi_i},\qquad
b_i=-\chi_i e^{\varphi_i},\qquad c_i=e^{\varphi_i}, \quad i=1, \ldots , 7 \; , 
\end{equation}
with no sum over $i$, so that $a_ic_i-b_i^2=1 $ for all $i=1, \ldots, 7$.

The bosonic dynamics is governed by the Lagrangian
\begin{equation} \label{eq:Lagrangian}
	\cL=R\,\vol_4+ \tfrac12 (d\varphi_i)^2+\tfrac12 e^{2\varphi_i} (d\chi_i)^2 -V\,\vol_4\;.
\end{equation}
Here, $(d\varphi_i)^2 \equiv d\varphi_i \wedge * d\varphi_i = -(\partial_\mu \varphi_i) (\partial^\mu \varphi_i) \, \textrm{vol}_4$, etc., with sum over $i$.  The kinetic terms in (\ref{eq:Lagrangian}) can also be written as $2  g_{i \bar{\jmath}} \, dt^i \wedge * d\bar{t}^{\bar{\jmath}}$, in terms of the chirals (\ref{eq:7ComplelsScal}) and a metric $ g_{i \bar{\jmath}} = \partial_i \partial_{\bar{\jmath}}  K$ obtained from the K\"ahler potential 
\begin{equation} \label{eq:KahlerPot}
K = -\log[ -i(t_1 - \bar{t}_{1} )] - \ldots -\log[ -i(t_7 - \bar{t}_{7} )]   \; .
\end{equation}
The potential $V$ in (\ref{eq:Lagrangian}) derives from the holomorphic superpotential
\begin{equation}
\label{SuperPot}
\begin{aligned}
W = 2g \bigl[\, &
t_{1}t_{2}t_{3} + t_{1}t_{6}t_{7} + t_{2}t_{5}t_{7}
+ t_{3}t_{5}t_{6} + t_{1}t_{4}t_{5} + t_{2}t_{4}t_{6}
+ t_{3}t_{4}t_{7} \\
&+ t_{1}t_{2}t_{3}t_{4}t_{5}t_{6}t_{7}
\,\bigr] \, ,
\end{aligned}
\end{equation}
via the standard ${\cal N}=1$ formula (adapted to our normalisation of the potential by $1/4$)
\begin{equation} \label{eq:PotFromSuperPot}
\tfrac14 V =e^K\left(g^{i\bar\jmath}D_iW D_{\bar\jmath}\overline W
-3W\overline W\right), \qquad 
D_iW \equiv\partial_iW+(\partial_iK)W.
\end{equation}
Finally, the constant $g$ in (\ref{SuperPot}) is the gauge coupling of the parent SO(8)-gauged supergravity. 

As noted in \cite{Bobev:2019dik}, the seven cubic terms contained in the superpotential (\ref{SuperPot}) correspond to the seven lines in the Fano plane. The form of the superpotential depends on the parametrisation used for the seven chirals. For example, an $\textrm{SO}(2)^7 \subset \textrm{SL}(2)^7$ duality transformation $\hat{t}_i = -t_i^{-1}$, $i=1, \ldots, 7$, makes the resulting superpotential contain all quartic and zeroth-order terms complementary to the terms in (\ref{SuperPot}), up to an overall prefactor $-(\hat{t}_1 \ldots \hat{t}_7)^{-1}$. Similarly, in the unit-disk parametrisation, the superpotential exhibits all seven cubic and quartic terms and one constant term and one seventh-order term: see (7.5) of \cite{Bobev:2019dik} with suitable relabellings. It is also interesting to note that, dropping the septic term, the superpotential (\ref{SuperPot}) describes the analogous seven-chiral truncation of purely electric ($g\neq 0$, $m=0$) ISO(7) supergravity \cite{Hull:1984yy,DallAgata:2011aa,Guarino:2015qaa}: see (A.9) of \cite{Guarino:2019snw}. Further comments about the Fano organisation of the seven chiral subsector of maximal $D=4$ supergravity can be found in \cite{Duff:2006ue,Levay:2006pt,Duff:2010vy,Ferrara:2016fwe,Gunaydin:2020ric}, see also \cite{Baez:2001dm}.

Certain identifications among the seven complex scalars (\ref{eq:7ComplelsScal}) cause the model (\ref{eq:Lagrangian})--(\ref{eq:PotFromSuperPot}) to preserve continuous symmetry groups contained in SO(8) when considered within the full $\cN=8$ theory. In this paper we are interested in two such symmetry enhancements, to the SO(3) and U$(1)^2$  embedded into SO(8) via the common chain
\begin{equation} \label{eq:branching}
\textrm{SO}(8) \supset \textrm{SU}(4)_s \times \textrm{U}(1) \supset \textrm{SO}(4)_s   \supset \left\{
\begin{array}{l}
\textrm{SO}(3) \\[5pt]
\textrm{U}(1)^2
 \end{array} \right. \; ,
\end{equation}
in conventions where the $\cN=8$ scalars and pseudoscalars transform, respectively, in the ${\bf{35}}_v$ and ${\bf{35}}_c$ of SO(8). Namely, SO(3) and U$(1)^2$ are respectively the diagonal and Cartan subgroups of $\textrm{SO}(4)_s$. Here, $\textrm{SO}(4)_s$ is the real subgroup of $\textrm{SU}(4)_s$, which is in turn embedded into SO(8) such that ${\bf{8}}_s \rightarrow \bf{6}+\bf{1}+\bf{1}$, ${\bf{8}}_{v,c} \rightarrow \bf{4}+\overline{\bf{4}}$. The same SO(3) is also the real subgroup of SU(3) along the chain $\textrm{SO}(8) \supset \textrm{SU}(4)_c \supset \textrm{SU}(3)\times\textrm{U}(1)_c \supset \textrm{SU}(3) \supset \textrm{SO}(3)$, where $\textrm{SU}(3)\times\textrm{U}(1)_c$ is the symmetry group of the CPW solution in the conventions of \cite{Larios:2019kbw}. The U$(1)^2$ in (\ref{eq:branching}) is not contained in this CPW symmetry group. Specifically, the identifications 
\begin{eqnarray}
\label{eq:SO3Identifications}
\textrm{SO}(3) & : & t_1 = t_2 = t_3 \equiv \tilde{t}_1 \; , \quad t_4 \equiv \tilde{t}_2 \;, \quad t_5 = t_6 = t_7 \equiv \tilde{t}_3 \; , \\
\label{eq:U1SqIdentifications}
\textrm{U}(1)^2 & : & t_1 =  t_3 = i \; , \quad t_2 \equiv t^\prime_0  \; , \quad -t_4^{-1} = t_6 \equiv t^\prime_1 \; , \quad t_5 =  t_7 \equiv t^\prime_2 \; ,
\end{eqnarray}
respectively enhance the symmetry of the model to the rightmost groups in (\ref{eq:branching}). In fact, equations (\ref{eq:SO3Identifications}) and (\ref{eq:U1SqIdentifications}) define further consistent truncations of the seven-chiral model (\ref{eq:Lagrangian}) to the distinct $\cN=1$ three-chiral models, respectively containing $\tilde{t}_1$, $\tilde{t}_2$, $\tilde{t}_3$, and $t^\prime_0$, $t^\prime_1$, $t^\prime_2$, considered in \cite{Bobev:2019dik} and \cite{Fischbacher:2010ec}. The K\"ahler potential and superpotential of these three-chiral models are obtained by substituting (\ref{eq:SO3Identifications}) and (\ref{eq:U1SqIdentifications}) into (\ref{eq:KahlerPot}) and (\ref{SuperPot}). 

We will use a further bosonic truncation of the $\mathrm U(1)^2$-invariant three-chiral model \eqref{eq:U1SqIdentifications} to two real scalars. To describe it, first introduce disk coordinates and make the constant phase redefinitions
\begin{equation}
\zeta_a=\frac{t'_a-i}{t'_a+i},\qquad
\zeta_0=-iw_0,\qquad
\zeta_1=e^{i\pi/4}w_1,\qquad
\zeta_2=e^{-i\pi/4}w_2,
\quad a=0,1,2.
\label{eq:U12twoScalarDisks}
\end{equation}
Now, the identifications
\begin{equation} \label{eq:U12twoScalarFields}
w_0=\tanh\frac{\lambda_0}{\sqrt2},\qquad
w_1=w_2=\tanh\frac{\lambda}{2},
\end{equation}
with $\lambda_0$ and $\lambda$ real scalars, are consistent at the level of the equations of motion. From (\ref{eq:Lagrangian}), this two-scalar model is described by the Lagrangian
\begin{equation} \label{eq:U12twoScalar}
\mathcal L =R\,\vol_4+(d\lambda_0)^2+2(d\lambda)^2 - 2g^2\left[\cosh(\sqrt2\lambda_0)
       \bigl(\sinh^4\lambda-4\bigr)-8\cosh^2\lambda\right]  \textrm{vol}_4 \; , 
\end{equation}
where the squared differentials have the same meaning as in \eqref{eq:Lagrangian}. This sector coincides with that of \cite{Fischbacher:2010ec} with the further identification $\lambda_1 = \lambda_2 \equiv \lambda$ there.

All five known supersymmetric AdS vacua of $D=4$ $\cN=8$ SO(8)-gauged supergravity occur within the seven-chiral subsector (\ref{eq:Lagrangian}). They arise as solutions to the F-term conditions, $D_i W =0$, for the superpotential (\ref{SuperPot}) and, of course, also as extrema of the potential $V$ in (\ref{eq:PotFromSuperPot}). With our parametrisation, these critical points are located at
\begin{equation}
\label{Field_Identifications}
\begin{array}{lll}
 \cN=8, \; \textrm{SO}(8)  &  : & \hspace{5pt} \tilde{t}_1 = \tilde{t}_2 = \tilde{t}_3 = i \; , \\[10pt]
\cN=2, \; \textrm{SU}(3) \times \textrm{U}(1)  & : & \hspace{5pt}  \tilde{t}_1  = i \sqrt{3} \; , \quad \tilde{t}_2 = \tilde{t}_3 = 3^{-\frac12}\big(1+i\sqrt{2} \big) \; ,  \\[10pt]
\cN=1, \; \textrm{G}_2  & : & \hspace{5pt}   \tilde{t}_1 = \tilde{t}_2 = \tilde{t}_3 = 2^{-\frac12} \cdot 3^{-\frac14}\big(1+i\sqrt{5} \big)  \; ,  \\[10pt]
\cN=1, \; \textrm{SO}(3) & : & \hspace{5pt}  \tilde{t}_1 \approx -1.406521725320363 + i \, 0.862882995719612  \\[5pt]
  &  & \hspace{5pt}  \tilde{t}_2 \approx -0.328918695783914 + i \, 0.262641989532009  \\[5pt]
  &  & \hspace{5pt}  \tilde{t}_3 \approx 0.399054328284510 + i \, 1.340544835530880 \; ,  \\[10pt]
\cN=1, \; \textrm{U}(1)^2    & : & \hspace{2mm}  t_0^\prime =  \tfrac12 \big(i+ \sqrt{3} \big)  \; , \quad   t_2^\prime =   -\bar{t}_1^\prime = \tfrac12 \big(1+ \sqrt{3} \big)\big(1+ i \big) \; .
\end{array}
\end{equation}
The SO(3)-invariant point is given numerically here; the other four points are displayed in closed form. An exact algebraic characterisation of the SO(3) point is given in \cite{Bobev:2019dik}. In (\ref{Field_Identifications}), a bar denotes complex conjugation, and we have employed the notation introduced in (\ref{eq:SO3Identifications}), (\ref{eq:U1SqIdentifications}). The corresponding cosmological constants $V_0$, obtained by evaluating (\ref{eq:PotFromSuperPot}) on (\ref{Field_Identifications}), are summarised in table~\ref{tab:AllSusyAdS4}.

The two-scalar model (\ref{eq:U12twoScalar}) retains the maximally supersymmetric vacuum and the $\cN=1$ U$(1)^2$-invariant vacuum. For later reference, we note that these occur at
\begin{equation}
\label{Field_IdsTwoScal}
\begin{array}{lll}
 \cN=8, \; \textrm{SO}(8)  &  : & \hspace{5pt} \lambda_0 = \lambda = 0  \; , \\[10pt]
\cN=1, \; \textrm{U}(1)^2    & : & \hspace{2mm}  \lambda_0 =  
\frac{1}{\sqrt2} \operatorname{arcosh}2   \; , \quad   
\lambda =  \operatorname{arcosh}\sqrt3 \; .
\end{array}
\end{equation}
%


\section{The seven-chiral sector in $D=11$} \label{sec:7CS11D}



The solutions of maximal four-dimensional SO(8)-gauged supergravity give rise to solutions of eleven-dimensional supergravity upon consistent uplift. In particular, the supersymmetric vacua (\ref{Field_Identifications}) of the $D=4$ supergravity uplift to supersymmetric AdS$_4 \times S^7$ configurations. The corresponding $D=11$ bosonic fields can be expressed as 
\begin{equation}	\label{eq: upliftansatz}
		d\hat{s}^2_{11} = - 6V_0^{-1} \Delta^{-1} ds^2({\rm AdS_4}) + ds_7^2	\; , \qquad 
		\hat{F}_4=-9 g^{-1} V_0^{-1}  \, \vol ({\rm AdS_4}) + dA_3\, .
\end{equation}
Here, $V_0$ is the value of the scalar potential (\ref{eq:PotFromSuperPot}) at the relevant $D=4$ vacuum, and AdS$_4$ has unit radius, so that $R_{\mu\nu}=-3g_{\mu\nu}$. The warp factor $\Delta$, the metric $ds_7^2$ and the three-form $A_3$ all depend on the  $D=4$ scalar values at the vacuum and on the embedding coordinates $\mu^I$ that define the $S^7$ as the locus (\ref{eq: S7inR8}).

Here we give explicit expressions for $\Delta$, $ds_7^2$ and $A_3$ in the uplift (\ref{eq: upliftansatz}), for generic values of the $D=4$ scalars of the seven-chiral model reviewed in section~\ref{sec:7CS4D}. To obtain them, we specialise the general expressions given in \cite{Varela:2015ywx} and summarised in appendix~\ref{app:conventions}, valid for generic fields of $D=4$ $\cN=8$ SO$(8)$ supergravity, to the subsector at hand. After substantial manipulation, these expressions take the compact form
\begin{equation} \label{eq:UpliftedExpr}
\Delta=\left[-\det\begin{pmatrix}\HH&\mu\\\mu^{\T}&0\end{pmatrix}\right]^{-1/12} \; , \qquad
ds_7^2 = g^{-2}\Delta^2\dd\mu^{\T}\HH\,\dd\mu \; , \qquad
A_3=\frac{\Delta^3}{3g^3} \sum_{d=1}^{7}\alpha_d^{\T}\wedge U_d\beta_d \; .
\end{equation}
Here, $g$ is the $D=4$ gauge coupling, and the index $d$ runs over $1,\ldots,7$. The matrix ${\cal H}$ is $8\times8$, each $U_d$ is $4\times4$, and $\mu$ is the column vector with components $\mu^I$, $I=1,\ldots,8$. For each value of $d$, $\alpha_d$ and $\beta_d$ are four-component column vectors whose entries are, respectively, one-forms along $\d\mu^I$ and two-forms along $\d\mu^I\wedge\d\mu^J$. Matrix multiplication is implicit in (\ref{eq:UpliftedExpr}): all matrix indices are contracted in $\dd\mu^{\T}\HH\,\dd\mu$ and $\alpha_d^{\T}\wedge U_d\beta_d$, so both expressions are scalars with respect to these indices. The wedge symbol in the latter expression denotes the exterior product of the one-form entries of $\alpha_d$ with the corresponding two-form entries of $U_d\beta_d$. Finally, the matrix ${\cal H}$ itself admits the compact decomposition
\begin{equation} \label{eq:H}
\HH=\sum_{d=1}^{7}Q_d^{\T}T_dQ_d+\frac14DS^{\T}PSD \; ,
\end{equation}
in terms of the matrices $Q_d$, $T_d$, $D$, $S$ and $P$. Together with $\mu$ and the matrices $U_d$ introduced above, these constitute the basic building blocks of the uplift (\ref{eq:UpliftedExpr}). For convenience, table~\ref{tab:uplift-building-blocks} collects these building blocks and provides references to the equations in appendix~\ref{sec:seven-chiral-uplift-blocks} where their explicit expressions are given. The table also records the dependence of each building block, indicating at a glance whether it depends only on the $S^7$ angles, only on the $D=4$ scalars, or on neither. The quantities ${\cal H}$, $\alpha_d$ and $\beta_d$ are derived quantities: they can be constructed from the building blocks listed in table~\ref{tab:uplift-building-blocks} and are therefore not included as separate entries in that table.

\begin{table}
\centering
\captionsetup{font=footnotesize,position=bottom}
\small
\renewcommand{\arraystretch}{1.2}
\setlength{\tabcolsep}{5pt}
\begin{tabularx}{\linewidth}{
  @{}
  >{\raggedright\arraybackslash}p{0.18\linewidth}
  c
  c
  >{\raggedright\arraybackslash}X
  >{\raggedright\arraybackslash}p{0.23\linewidth}
  @{}
}
\toprule
Dependence
& Quantity
& Dimensions
& Description
& Definition \\
\midrule
$S^7$ only
& $\mu$ & $8\times1$
& Column of embedding coordinates
& \eqref{eq:sphere} \\

& $D$ & $8\times8$
& Diagonal matrix
& \eqref{eq:sphere} \\

& $Q_d$ & $8\times8$
& Sparse coordinate matrix
& \eqref{eq:Qd} \\
\midrule
$D=4$ scalars only
& $T_d$ & $8\times8$
& Symmetric matrix
& \eqref{eq:Td} \\

& $U_d$ & $4\times4$
& Generally nonsymmetric matrix
& \eqref{eq:Ud} \\

& $P$ & $7\times7$
& Diagonal matrix
& \eqref{eq:P} \\
\midrule
Neither (constant)
& $S$ & $7\times8$
& Matrix with entries $\pm1$
& \eqref{eq:S} \\
\bottomrule
\end{tabularx}

\caption{Basic building blocks of the seven-chiral uplift (\ref{eq:UpliftedExpr}), organised by their dependence on the $S^7$ coordinates and the $D=4$ scalars. The block index runs over $d=1,\ldots,7$.}
\label{tab:uplift-building-blocks}
\end{table}


\section{The $\mathrm{U}(1)^2$-invariant $D=11$ solution}
\label{sec:U12Sol11D}

We now specialise the uplift \eqref{eq: upliftansatz},
\eqref{eq:UpliftedExpr} to the two-real-scalar sector
\eqref{eq:U12twoScalarDisks}--\eqref{eq:U12twoScalar},
parametrised by $\lambda,\lambda_0$. To this end, introduce
$S^7$ coordinates
$(\xi,\theta,\widetilde\theta,\phi,\widetilde\phi,\widetilde\psi,\psi)$
through
\begin{equation}
\begin{aligned}
-\mu^1+i\mu^4
&  =\frac{\cos\xi}{\sqrt2}\,e^{i(\psi+\widetilde\psi/4)}
 \left(\cos\frac\theta2 e^{-i\phi/2}
       +\sin\frac\theta2 e^{i\phi/2}\right),\\
 -\mu^2-i\mu^6
&  =\frac{\sin\xi}{\sqrt2}\,e^{i(\psi-\widetilde\psi/4)}
 \left(\cos\frac{\widetilde\theta}2 e^{-i\widetilde\phi/2}
       +\sin\frac{\widetilde\theta}2 e^{i\widetilde\phi/2}\right),\\
  \mu^8+i\mu^5
&  =\frac{i\cos\xi}{\sqrt2}\,e^{i(\psi+\widetilde\psi/4)}
 \left(\cos\frac\theta2 e^{-i\phi/2}
       -\sin\frac\theta2 e^{i\phi/2}\right),\\
 \mu^3+i\mu^7
&  =\frac{i\sin\xi}{\sqrt2}\,e^{i(\psi-\widetilde\psi/4)}
 \left(\cos\frac{\widetilde\theta}2 e^{-i\widetilde\phi/2}
       -\sin\frac{\widetilde\theta}2 e^{i\widetilde\phi/2}\right) ,
\end{aligned}
\label{eq:U12CLPcoordinates}
\end{equation}
subject to \eqref{eq: S7inR8}. This parametrisation is adapted to the Hopf fibration of 
$S^7$ over $\mathbb{CP}^3$, where $\psi$ parametrises the U(1) fibre, while $(\xi,\theta,\widetilde\theta,\phi,\widetilde\phi,\widetilde\psi)$
are local coordinates on the $\mathbb{CP}^3$ base. Indeed, using (\ref{eq:U12CLPcoordinates}), the standard unit-radius round metric (\ref{eq:RoundS7Met}) on $S^7$ gives \eqref{eq:introRoundHopf}, with the Fubini--Study metric
\eqref{eq:U12FSmetric} and Hopf connection
\eqref{eq:U12Hopfconnection}.

Let us introduce the following real functions of the $D=4$ scalar $\lambda$
and the $\mathbb{CP}^3$ angles $(\theta,\widetilde\theta,\phi,\widetilde\phi)$:
\begin{equation}
\begin{aligned}
X&=\cosh\lambda-\frac{\sinh\lambda}{\sqrt2}\sin\theta\cos\phi,
&\quad Y&=\frac{\sinh\lambda}{\sqrt2}\sin\theta\sin\phi,\\
\widetilde X&=\cosh\lambda-\frac{\sinh\lambda}{\sqrt2}
 \sin\widetilde\theta\cos\widetilde\phi,
&\quad \widetilde Y&=\frac{\sinh\lambda}{\sqrt2}
 \sin\widetilde\theta\sin\widetilde\phi.
\end{aligned}
\label{eq:U12XY}
\end{equation}
We further define the following functions involving (\ref{eq:U12XY}), both $D=4$ scalars $\lambda$, $\lambda_0$, and the $\mathbb{CP}^3$ angle $\xi$:
\begingroup
\thinmuskip=2mu
\medmuskip=3mu
\begin{align}
\Theta={}&(X^2+Y^2)\cos^2\xi
+\cosh(\sqrt2\lambda_0)(X\widetilde X+Y\widetilde Y)\sin^2\xi,
\nonumber\\*[0.6ex]
\widetilde\Theta={}&(\widetilde X^2+\widetilde Y^2)\sin^2\xi
+\cosh(\sqrt2\lambda_0)(X\widetilde X+Y\widetilde Y)\cos^2\xi,
\nonumber\\*[0.6ex]
P={}&(1+\tfrac12\sinh^2\lambda)
 (1+\tfrac12\sinh^2\lambda\cos^2\theta)\cos^2\xi
+\cosh(\sqrt2\lambda_0)
 (2\cosh\lambda\,\widetilde X-X\widetilde X-Y\widetilde Y)\sin^2\xi,
\nonumber\\*[0.6ex]
Q={}&(1+\tfrac12\sinh^2\lambda)
 (1+\tfrac12\sinh^2\lambda\cos^2\widetilde\theta)\sin^2\xi
+\cosh(\sqrt2\lambda_0)
 (2\cosh\lambda\,X-X\widetilde X-Y\widetilde Y)\cos^2\xi,
\nonumber\\*[0.6ex]
R={}&\tfrac14\sinh^4\lambda\cos\theta\cos\widetilde\theta
+\sinh(\sqrt2\lambda_0)\cosh\lambda(\widetilde Y-Y).
\label{eq:UpliftQuantities}
\end{align}
\endgroup
Finally, we also introduce the following one-forms on $\mathbb{CP}^3$, with $\lambda$-dependent coefficients:
\begin{equation}
\begin{aligned}
\omega^{1}={}&\dd\theta-
\frac{\sqrt2\,\sinh\lambda\tan\xi\cos\theta
 (\widetilde X\cos\phi-\widetilde Y\sin\phi)}
 {X\widetilde X+Y\widetilde Y}\,\dd\xi,\\
\omega^{2}={}&\sin\theta\,\dd\phi+
\frac{\sqrt2\,\sinh\lambda\tan\xi
 (\widetilde X\sin\phi+\widetilde Y\cos\phi)}
 {X\widetilde X+Y\widetilde Y}\,\dd\xi,\\
\omega^{3}={}&\dd\widetilde\theta+
\frac{\sqrt2\,\sinh\lambda\cot\xi\cos\widetilde\theta
 (X\cos\widetilde\phi-Y\sin\widetilde\phi)}
 {X\widetilde X+Y\widetilde Y}\,\dd\xi,\\
\omega^{4}={}&\sin\widetilde\theta\,\dd\widetilde\phi-
\frac{\sqrt2\,\sinh\lambda\cot\xi
 (X\sin\widetilde\phi+Y\cos\widetilde\phi)}
 {X\widetilde X+Y\widetilde Y}\,\dd\xi.
\end{aligned}
\label{eq:U12omega}
\end{equation}
All quantities (\ref{eq:U12XY})--(\ref{eq:U12omega}) are independent of the Hopf angle $\psi$ and the $\mathbb{CP}^3$ angle $\widetilde\psi$. 

With these definitions, the quantities (\ref{eq:UpliftedExpr}) can be cast in abelian KK form as follows. For the warp factor, we obtain
\begingroup
\thinmuskip=2mu
\medmuskip=3mu
\begin{align}
\Delta^{-3}={}&\frac12\cosh(\sqrt2\lambda_0)
 \left(\Theta\cos^2\xi+\widetilde\Theta\sin^2\xi\right)
+\cosh^2\lambda(X\widetilde X-Y\widetilde Y)
 \cos^2\xi\sin^2\xi\nonumber\\*
&+\frac12\cosh(\sqrt2\lambda_0)\cosh^2\lambda
 \left[(X^2-Y^2)\cos^4\xi
       +(\widetilde X^2-\widetilde Y^2)\sin^4\xi\right] \; ,
\label{eq:U12CLPwarp}
\end{align}
\endgroup
the metric on $S^7$ can be written as (\ref{eq:introU12KK}),
\begin{equation}
\dd s_{7}^2=
g^{-2} \Delta^2
\left[\dd s_6^2+f(\dd\psi+\mathcal A)^2\right] ,
\label{eq:U12CLPmetric}
\end{equation}
and the internal three-form as
\begin{equation}
A_3=g^{-3}\Delta^3\Big[
C_3+(\dd\psi+\mathcal A)\wedge C_2
 +(\dd\widetilde\psi+\widetilde{\mathcal A})\wedge\widetilde C_2
 +(\dd\psi+\mathcal A)\wedge
 (\dd\widetilde\psi+\widetilde{\mathcal A})\wedge C_1\Big] \; .
\label{eq:U12FluxDecompositionMain}
\end{equation}
In (\ref{eq:U12CLPmetric}), $\psi$ still parametrises the $S^7$ Hopf fibre, while $f$, $\mathcal A$ and $\dd s_6^2$ are, respectively, a function, a connection one-form and a metric on $\mathbb{CP}^3$, all depending on the $D=4$ scalars. In general, they differ from their round-sphere values $f=1$, $\mathcal A=\mathcal A_{\mathrm{FS}}$ and $\dd s_6^2=\dd s_{\mathrm{FS}}^2$, with $\dd s_{\mathrm{FS}}^2$ and $\mathcal A_{\mathrm{FS}}$ given in \eqref{eq:U12FSmetric} and \eqref{eq:U12Hopfconnection}. Specifically, the fibre coefficient is
\begin{equation} \label{eq:MetricFun}
f=P\cos^2\xi+Q\sin^2\xi+2R\cos^2\xi\sin^2\xi,
\end{equation}
with $P,Q,R$ defined in \eqref{eq:UpliftQuantities}.
The deformed Hopf connection is
\begin{align}
\mathcal A={}&-\frac14
 (\cos\theta\,\dd\phi+\cos\widetilde\theta\,\dd\widetilde\phi)
+\frac{P\cos^2\xi-Q\sin^2\xi}{4f}
 \bigl(\dd\widetilde\psi-\cos\theta\,\dd\phi
       +\cos\widetilde\theta\,\dd\widetilde\phi\bigr)
\nonumber\\*
&+\frac1{2f}(U_i\cos^2\xi+V_i\sin^2\xi)\omega^{i},
\label{eq:U12fluxConnection}
\end{align}
and the base metric reads
\begin{align}
\dd s_6^2={}&\frac{\Delta^{-3}}{X\widetilde X+Y\widetilde Y}\,\dd\xi^2
+\frac14 \,\Theta \cos^2\xi \bigl[(\omega^{1})^2+(\omega^{2})^2\bigr]
+\frac14 \,\widetilde\Theta \sin^2\xi
 \bigl[(\omega^{3})^2+(\omega^{4})^2\bigr]
\notag\\*
& +\frac14M_{ij}\omega^{i}\omega^{j} +\frac{
 (PQ- R^2 \cos^2\xi\sin^2\xi) }{4f} \cos^2\xi\sin^2\xi 
 (\dd\widetilde\psi+\widetilde{\mathcal A})^2.
\label{eq:U12sixmetric}
\end{align}
Here $\omega^{i}$ are the one-forms \eqref{eq:U12omega}, and repeated
indices $i,j=1,\ldots,4$ are summed. The connection $\widetilde{\mathcal A}$ that appears in (\ref{eq:U12sixmetric}) is given by
\begin{equation}
\widetilde{\mathcal A}
= -\cos\theta\,\dd\phi+\cos\widetilde\theta\,\dd\widetilde\phi
+\frac{\bigl[(Q+R\cos^2\xi)U_i-(P+R\sin^2\xi)V_i\bigr]\omega^{i}}
 {PQ-R^2\cos^2\xi\sin^2\xi}.
\label{eq:U12RelativeFibre}
\end{equation}
The functions $U_i,V_i,M_{ij}$ that appear in (\ref{eq:U12fluxConnection})--(\ref{eq:U12RelativeFibre}) and the coefficients of the differential forms 
$C_3,C_2,\widetilde C_2,C_1$ in (\ref{eq:U12FluxDecompositionMain}) 
all depend on the scalars $\lambda_0,\lambda$
and the $\mathbb{CP}^3$ angles
$(\xi,\theta,\widetilde\theta,\phi,\widetilde\phi)$.
The $C$-forms have the degrees indicated by their subscripts and have legs only along $\dd\xi$ and $\omega^{i}$.
Explicit expressions for these functions and forms are given in
Appendix~\ref{app:U12MetricFlux}. None of these quantities depend on the angles $\psi$, $\widetilde{\psi}$. Thus, the vectors $\partial_{\psi}$ and $\partial_{\widetilde{\psi}}$ are Killing vectors of the metric (\ref{eq:U12CLPmetric}) that also leave the warp factor (\ref{eq:U12CLPwarp}) and three-form (\ref{eq:U12FluxDecompositionMain}) invariant. Therefore, for generic values of the $D=4$ scalars $\lambda$, $\lambda_0$, the resulting $D=11$ configuration (\ref{eq: upliftansatz}) is manifestly U$(1)^2$-invariant.

At $\lambda_0=\lambda=0$, the two-scalar model (\ref{eq:U12twoScalar}) attains the SO(8)-invariant vacuum by (\ref{Field_IdsTwoScal}), and the $D=11$ configuration (\ref{eq: upliftansatz}) with (\ref{eq:U12CLPwarp})--(\ref{eq:U12FluxDecompositionMain}) becomes the usual Freund--Rubin AdS$_4 \times S^7$ solution, equipped with the round metric on $S^7$,  vanishing internal flux and no warping. Indeed, for $\lambda_0=\lambda=0$,  the functions (\ref{eq:U12XY}), (\ref{eq:UpliftQuantities}) reduce to
\begin{equation} \label{eq:U12roundLimit}
X=\widetilde X=\Theta=\widetilde\Theta
=P=Q=1 \; , \qquad 
Y=\widetilde Y=R=0 \; ,
\end{equation}
and the one-forms (\ref{eq:U12omega}) to 
\begin{equation}
\omega^{1} = \dd\theta \; , \qquad
\omega^{2} = \sin\theta\,\dd\phi  \; , \qquad 
\omega^{3} =  \dd\widetilde\theta \;  , \qquad 
\omega^{4} = \sin\widetilde\theta\,\dd\widetilde\phi \; . 
\end{equation}
The expressions of appendix~\ref{app:U12MetricFlux} also show that the functions $U_i$, $V_i$, $M_{ij}$ and the differential forms 
$C_3,C_2,\widetilde C_2,C_1$ all vanish. Consequently, (\ref{eq:MetricFun}) reduces to $f=1$, (\ref{eq:U12fluxConnection}) gives ${\cal A} = {\cal A}_{\textrm{FS}}$, with the latter connection given in (\ref{eq:U12Hopfconnection}), and (\ref{eq:U12sixmetric}) collapses to the Fubini--Study metric (\ref{eq:U12FSmetric}), with the connection $\widetilde{\mathcal A}$ in (\ref{eq:U12RelativeFibre}) reducing to its counterpart in (\ref{eq:U12FSmetric}). Thus, the $D=11$ warping (\ref{eq:U12CLPwarp}) trivialises, the three-form (\ref{eq:U12FluxDecompositionMain}) vanishes, and the seven-dimensional metric (\ref{eq:U12CLPmetric}) becomes $g^{-2}$ times the unit-radius round, SO(8)-invariant metric (\ref{eq:introRoundHopf}). 

Evaluating instead the uplift (\ref{eq: upliftansatz}) at the U$(1)^2$-invariant vacuum (\ref{Field_IdsTwoScal}) of the $D=4$ two-scalar model (\ref{eq:U12twoScalar}) gives a new $\cN=1$, U$(1)^2$-invariant solution of eleven-dimensional supergravity, satisfying all its field equations.


\section{The $\mathrm{SO}(3)$-invariant $D=11$ solution}
\label{sec:SO3Sol11D}

We next restrict the uplift \eqref{eq: upliftansatz},
\eqref{eq:UpliftedExpr} to the three-chiral sector defined by
\eqref{eq:SO3Identifications}. The independent complex scalars are
$t_1,t_4,t_5$, for which we retain the real combinations $a_i,b_i,c_i$
of \eqref{eq:scalars}, with $i=1,4,5$. The remaining combinations satisfy
$(a_2,b_2,c_2)=(a_3,b_3,c_3)=(a_1,b_1,c_1)$ and
$(a_6,b_6,c_6)=(a_7,b_7,c_7)=(a_5,b_5,c_5)$, in agreement with (\ref{eq:SO3Identifications}).

To make the SO(3) symmetry manifest, we use coordinates adapted to its
orbits on $S^7$. In terms of the unit-sphere embedding coordinates $\mu^I$ of \eqref{eq: S7inR8}, the SO(3) rotation
$R=R_z(\alpha)R_y(\beta)R_z(\gamma)$ acts simultaneously on the
triplets $(\mu^3,\mu^5,\mu^7)$ and $(\mu^2,\mu^4,\mu^6)$,
leaving $\mu^1$ and $\mu^8$ fixed. Here $R_y$ and $R_z$ are the
standard right-handed rotations about the indicated Cartesian axes.
The Euler angles $(\alpha,\beta,\gamma)$ parametrise the orbits,
while the angles $q^a=(\xi,\zeta,\psi,\tau)$, $a= 1 , \ldots, 4$, label the four transverse directions. These angles are independent of those used in section~\ref{sec:U12Sol11D}, despite the overlapping notation. Following the SO(3)-adapted construction of \cite{Cvetic:2015uga},
the embedding coordinates can be written as
\begin{equation}
\begin{aligned}
-\mu^1+i\mu^8
&=\sin\xi\,e^{i\psi},\\[1ex]
\mu^3+i\mu^2
&=\cos\xi\,e^{i\tau}
\big[\cos\alpha
 (i\cos\zeta\sin\beta-\sin\zeta\cos\beta\sin\gamma)
-\sin\alpha\sin\zeta\cos\gamma\big],\\[1ex]
\mu^5+i\mu^4
&=\cos\xi\,e^{i\tau}
\big[\sin\alpha
 (i\cos\zeta\sin\beta-\sin\zeta\cos\beta\sin\gamma)
+\cos\alpha\sin\zeta\cos\gamma\big],\\[1ex]
\mu^7+i\mu^6
&=\cos\xi\,e^{i\tau}
\big(\sin\zeta\sin\beta\sin\gamma+i\cos\zeta\cos\beta\big).
\end{aligned}
\label{eq:SO3Bergerslice}
\end{equation}
This parametrisation satisfies the sphere constraint \eqref{eq: S7inR8}. We also find it helpful to introduce the left-invariant one-forms along the SO(3) orbits, 
\begin{equation}
\sigma^{1}=\sin\gamma\,\dd\beta-\cos\gamma\sin\beta\,\dd\alpha,\quad
\sigma^{2}=\cos\gamma\,\dd\beta+\sin\gamma\sin\beta\,\dd\alpha,\quad
\sigma^{3}=\dd\gamma+\cos\beta\,\dd\alpha ,
\label{eq:SO3BergerMC}
\end{equation}
such that 
\begin{equation}
\dd\sigma^{i}=-\tfrac12\epsilon^{i}{}_{jk}\sigma^{j}\wedge\sigma^{k} \; , 
\end{equation}
with $\epsilon^{i}{}_{jk}=\delta^{i\ell}\epsilon_{\ell jk}$,
$\epsilon_{123}=1$, and summation over repeated upper and lower indices. Finally, 
%
let us introduce the following real functions involving the $D=4$ scalars $a_1$, $c_1$, $a_5$, $c_5$ and the four transverse angles $q^a=(\xi,\zeta,\psi,\tau)$:
\begin{equation}
\begin{aligned}
X={}&a_1c_5\sin^2\xi\cos^2\psi
+\frac{c_1a_5}{2}\cos^2\xi(1-\cos2\zeta\cos2\tau),\\
Y={}&a_1a_5\sin^2\xi\sin^2\psi
+\frac{c_1c_5}{2}\cos^2\xi(1+\cos2\zeta\cos2\tau),\\
Z={}&\frac12\left(a_1\sin^2\xi\sin2\psi
-c_1\cos^2\xi\cos2\zeta\sin2\tau\right).
\end{aligned}
\label{eq:SO3XYZ}
\end{equation}

With these definitions, the uplift formulae (\ref{eq:UpliftedExpr}) take the following non-abelian KK form. The warp factor becomes
\begin{equation}
\Delta^{-3}
=a_1\left[c_4c_5X^2+a_4a_5Y^2+2(1+b_4b_5)XY\right]
+\frac{(b_4-b_5)^2}{c_1}
\left[XY+(b_1^2-b_5^2)Z^2\right].
\label{eq:SO3Bergerwarp}
\end{equation}
The internal metric can be cast in the form (\ref{eq:introSO3KK}),
\begin{equation}
\dd s_7^2
=g^{-2}\Delta^2\left[h_{ab}\,\dd q^a\dd q^b
+\gamma_{ij}
\bigl(\sigma^i+\mathcal A^i\bigr)
\bigl(\sigma^j+\mathcal A^j\bigr)\right],
\qquad
\mathcal A^i=\mathcal A^i{}_a\,\dd q^a .
\label{eq:introBerger}
\end{equation}
Here $h_{ab}$ and $\gamma_{ij}$ are the base and orbit metrics of the dimensionless line element $g^2\Delta^{-2}\dd s_7^2$, and $\mathcal A^i$ describes the coupling of the SO(3) orbits to the base. Finally, the 
three-form can be written
\begin{align}
A_3={}&g^{-3}\Delta^3\Big[C_3+(\sigma^i+\mathcal A^i)\wedge C_{2i}
 +\frac12\epsilon_{ijk}(\sigma^i+\mathcal A^i)\wedge
(\sigma^j+\mathcal A^j)\wedge C_1{}^k
\nonumber\\
&\hspace{15mm}+\frac16C_0\epsilon_{ijk}(\sigma^i+\mathcal A^i)\wedge
(\sigma^j+\mathcal A^j)\wedge(\sigma^k+\mathcal A^k)\Big].
\label{eq:SO3FluxDecompositionMain}
\end{align}
Here $C_0$ is a function, while $C_1{}^i$, $C_{2i}$ and $C_3$ are forms
of the indicated degrees with legs only along $\dd q^a$. 
Their coefficients, together with $h_{ab}$, $\gamma_{ij}$ and
$\mathcal A^i{}_a$, depend only on the SO(3)-invariant scalars
(\ref{eq:SO3Identifications}) and the transverse angles
$q^a=(\xi,\zeta,\psi,\tau)$. Their explicit expressions are given in
appendix~\ref{app:SO3Details}.
Since the one-forms $\sigma^i$ are SO(3)-left-invariant, the $D=11$
configuration (\ref{eq: upliftansatz}) with
(\ref{eq:SO3XYZ})--(\ref{eq:SO3FluxDecompositionMain}) is manifestly
SO(3)-invariant.

Evaluating the uplift (\ref{eq: upliftansatz}) with
(\ref{eq:SO3XYZ})--(\ref{eq:SO3FluxDecompositionMain}) at the
SO(3)-invariant vacuum (\ref{Field_Identifications}) gives our new
$\cN=1$, SO(3)-invariant AdS$_4$ solution of $D=11$ supergravity.
The round, G$_2$ and CPW solutions also belong to this six-scalar
sector and are recovered in appendix~\ref{app:SO3Details}.


\section{Conclusions}


We have developed an overarching treatment of all known supersymmetric
AdS$_4$ vacua of SO(8)-gauged supergravity by working out the
eleven-dimensional uplift of the seven-chiral sector. This $\mathcal N=1$
model contains all these vacua within a common discrete truncation
\cite{Bobev:2019dik}. The resulting expressions give the warp factor,
internal metric and internal three-form potential, displaying the Fano
organisation of their scalar dependence. Specialising these expressions
gives the two previously unavailable uplifts with $\mathcal N=1$
supersymmetry and SO(3) or U(1)$^2$ symmetry. Both internal metrics
admit a KK-type description. The U(1)$^2$ solution takes an abelian
KK form on the global Hopf circle bundle over $\mathbb{CP}^3$, while
the SO(3) solution takes a local non-abelian KK form with the
three-dimensional SO(3) orbits as fibres. We should point out that
the latter orbit description differs from the quaternionic Hopf
fibration of $S^7$ over $S^4$. The latter is a genuine, globally defined bundle with
$S^3$ fibres, generated by a free $\mathrm{Sp}(1)$ action, and
rescaling their length relative to the base gives the familiar
homogeneous squashed seven-sphere \cite{Awada:1983Squashed}. Its global fibre
action is different from the local SO(3) action relevant here. The
corresponding squashed AdS vacuum lies outside the maximal $D=4$
$\mathcal N=8$ SO(8) truncation.

These results arrive at a time when much can already be learned about the holography of an AdS background of string/M-theory without knowing its explicit higher-dimensional form. KK spectroscopy provides a striking example: the early studies of the round-sphere \cite{Englert:1983rn,Sezgin:1983ik,Biran:1983iy} and CPW \cite{Nicolai:1985hs,Klebanov:2008vq} spectra have since been complemented by general spectral methods derived from Exceptional Field Theory \cite{Malek:2019eaz,Malek:2020yue,Varela:2020Spectra,Cesaro:2020soq}. In the programme developed in the latter references, the mass matrices for the full KK towers follow from the four-dimensional gauging and the scalar expectation values at the vacuum, together with a small amount of uplift information. For the $S^7$ backgrounds considered here, the latter is encoded in the SO(8) generators acting on the infinite-dimensional space of round-sphere scalar harmonics. This representation is fixed once and for all, independently of the vacuum. The resulting matrices can then be diagonalised level by level, without solving differential equations on the deformed internal space. Indeed, the KK spectra of our two new backgrounds were obtained in this way \cite{Cesaro:2020soq,Cesaro:2021KK}, before the explicit uplifts presented here became available.

Equivariant localisation provides another route to extracting physical
information without first constructing an explicit higher-dimensional solution
\cite{BenettiGenolini:2023kxp,BenettiGenolini:2023AdSCFT}. For AdS$_4$ backgrounds, this approach
has been developed in settings with at least $\mathcal N=2$ supersymmetry,
where the U(1) R-symmetry of an $\mathcal N=2$ subalgebra supplies the
Killing vector used in localisation. Flux quantisation conditions and
holographic observables can then be expressed in terms of topological
data and the fixed-point data of this symmetry. Applications include
free energies and the dimensions of operators described by wrapped
branes, even for backgrounds whose metrics are not known. The construction can be carried out directly
in the higher-dimensional theory, with no requirement that the
backgrounds belong to a chosen consistent truncation
\cite{Couzens:2026TenDimensional}. Its scope thus extends beyond the
vacua accessible through a particular lower-dimensional gauged
supergravity. These developments make it possible to address questions
about whole families of backgrounds before solving their local field
equations. Our two new $\mathcal N=1$ AdS$_4$ vacua have no continuous
R-symmetry, so these localisation formulae do not apply to them directly.

Explicit geometries thus remain useful for solutions beyond the scope of equivariant localisation methods, and for applications that directly
probe the local metric and fluxes. For example, recent semiclassical M2-brane calculations illustrate this directly: these fields enter the determination of brane embeddings and their quantum fluctuations \cite{Giombi:2023M2,Gautason:2025M2}. They remain useful even when localisation ultimately reduces the answer to a much smaller set of data. Another example is provided by brane-jet tests of nonsupersymmetric AdS backgrounds \cite{Bena:2020BraneJets,Guarino:2020BraneJets}. There, the competition between brane tension and flux coupling can vary over the internal space and reveal repulsive regions that are not visible in the four-dimensional vacuum analysis. The simplest tests require only the warp factor and the relevant flux coupling, together with the dilaton in type IIA; wrapped probes also explore the internal metric. Such tests complement the information about a background that can be obtained from its perturbative spectrum. Our expressions provide a starting point for analogous studies of brane dynamics and for investigating possible membrane instantons in the less symmetric supersymmetric backgrounds.

A natural continuation would be to develop similarly explicit geometric descriptions in other consistent reductions. The massive IIA setting of \cite{Guarino:2015jca,Guarino:2015vca} is particularly close to the present one: the complete embedding of dyonic ISO(7) supergravity \cite{Guarino:2015qaa} is known \cite{Guarino:2015vca}, and its SU(3)-invariant sector already gives a useful collection of explicit AdS$_4$ geometries \cite{Varela:2015uca}. Extending this description to the less symmetric vacua of \cite{Bobev:2020qev} would allow their internal geometry and fluxes to be studied in the same spirit as this paper.

\section*{Acknowledgements}


We are grateful to Mattia Ces\`aro and Gabriel Larios for collaboration during the early stages of this work. KP was supported in part by DOE grant DE-SC0011687, and OV by NSF grant PHY-2609785. We would like to thank each other's institutions for hospitality while part of this work was completed. We also thank OpenAI ChatGPT and Wolfram Mathematica for assistance in deriving and crosschecking our results.

\appendix

\addtocontents{toc}{\setcounter{tocdepth}{1}}



\section{Conventions}
\label{app:conventions}

We follow the conventions of \cite{Varela:2015ywx} for the $D=4$ \cite{deWit:1981sst,deWit:1982ig} and $D=11$ \cite{ Cremmer:1978km} supergravities. This appendix summarises some relevant four- and eleven-dimensional quantities.

\subsection{Four-dimensional conventions}
\label{app:D4Conventions}

We use the $\mathrm{SL}(8,\mathbb R)$ basis of $\mathrm{E}_{7(7)}$ given in appendix~C of
\cite{Guarino:2015qaa}. The generators in the $\mathbf{56}$ are the
traceless $t_I{}^J$ and the totally antisymmetric $t_{IJKL}$ of
(C.3) and (C.4) therein, with $I,J,\ldots=1,\ldots,8$.
For the seven-chiral sector we follow appendix~A of
\cite{Guarino:2019snw}. The Cartan generators $H_i$ of the seven commuting
$\mathrm{SL}(2,\mathbb R)$ factors are
\begin{equation}
\label{eq:appSevenGenerators}
\begin{aligned}
 H_1&=-t_1{}^1 -t_2{}^2 -t_3{}^3 +t_4{}^4 +t_5{}^5 +t_6{}^6 +t_7{}^7 -t_8{}^8 \; , \\
 H_2&=-t_1{}^1 +t_2{}^2 +t_3{}^3 -t_4{}^4 -t_5{}^5 +t_6{}^6 +t_7{}^7 -t_8{}^8 \; , \\
 H_3&=-t_1{}^1 +t_2{}^2 +t_3{}^3 +t_4{}^4 +t_5{}^5 -t_6{}^6 -t_7{}^7 -t_8{}^8 \; , \\
 H_4&=t_1{}^1 -t_2{}^2 +t_3{}^3 -t_4{}^4 +t_5{}^5 -t_6{}^6 +t_7{}^7 -t_8{}^8 \; , \\
 H_5&=t_1{}^1 -t_2{}^2 +t_3{}^3 +t_4{}^4 -t_5{}^5 +t_6{}^6 -t_7{}^7 -t_8{}^8 \; , \\
 H_6&=t_1{}^1 +t_2{}^2 -t_3{}^3 -t_4{}^4 +t_5{}^5 +t_6{}^6 -t_7{}^7 -t_8{}^8 \; , \\
 H_7&=t_1{}^1 +t_2{}^2 -t_3{}^3 +t_4{}^4 -t_5{}^5 -t_6{}^6 +t_7{}^7 -t_8{}^8 \; .
\end{aligned}
\end{equation}
The corresponding nilpotent generators $E_i$ and $F_i$ are
\begin{equation}
\label{eq:appNilpotentGenerators}
\begin{alignedat}{4}
 E_1 &= t_{1238} \; , &\qquad E_2 &= t_{1458} \; , &\qquad E_3 &= t_{1678} \; , &\qquad E_4 &= -t_{2468} \; , \\
 E_5 &= t_{2578} \; , &       E_6 &= t_{3478} \; , &       E_7 &= t_{3568} \; , \\[3pt]
 F_1 &= t_{4567} \; , &       F_2 &= t_{2367} \; , &       F_3 &= t_{2345} \; , &       F_4 &= -t_{1357} \; , \\
 F_5 &= t_{1346} \; , &       F_6 &= t_{1256} \; , &       F_7 &= t_{1247} \; .
\end{alignedat}
\end{equation}
With these normalisations, $H_i^{\T}=H_i$,
$E_i^{\T}=F_i$ and
\begin{equation}
 [H_i,E_j]=-4\delta_{ij}E_j \; , \qquad
 [H_i,F_j]=4\delta_{ij}F_j \; , \qquad
 [E_i,F_j]=-\tfrac{1}{288}\delta_{ij}H_j \; ,
\end{equation}
with no sum over $j$. In terms of the scalars in \eqref{eq:7ComplelsScal},
the coset representative and the scalar matrix are built out of the $H_i$ and $E_i$ generators in (\ref{eq:appSevenGenerators}), (\ref{eq:appNilpotentGenerators}) as
\begin{equation}
\label{eq:appSevenCoset}
\begin{aligned}
 \cV={}&e^{-12\chi_1E_1}e^{\frac14\varphi_1H_1}\,
        e^{-12\chi_2E_2}e^{\frac14\varphi_2H_2}\,
        e^{-12\chi_3E_3}e^{\frac14\varphi_3H_3}\,
        e^{-12\chi_4E_4}e^{\frac14\varphi_4H_4}\\
 &{}\times e^{-12\chi_5E_5}e^{\frac14\varphi_5H_5}\,
        e^{-12\chi_6E_6}e^{\frac14\varphi_6H_6}\,
        e^{-12\chi_7E_7}e^{\frac14\varphi_7H_7}
 \; , \qquad \cM=\cV\cV^{\T} \; .
\end{aligned}
\end{equation}
With the Einstein term normalised as in \eqref{eq:Lagrangian}, the scalar
kinetic terms are twice those in (A.7) of \cite{Guarino:2019snw}.
We denote the entries of $\cM$ and its inverse by $\cM_{MN}$ and
$\cM^{MN}$, respectively, with $M,N,\ldots=1,\ldots,56$.
In the electric and magnetic basis of antisymmetric pairs $[IJ]$,
$I<J$, the electric SO(8) embedding tensor \cite{deWit:2007kvg} has components
\begin{equation}
\label{eq:appSO8Embedding}
 \Theta_{[IJ]}{}^K{}_L=2\delta^K_{[I}\delta_{J]L} \; , \qquad
 \Theta^{[IJ]K}{}_L=0 \; , \qquad
 X_{MN}{}^P=\Theta_M{}^I{}_J(t_I{}^J)_N{}^P \; .
\end{equation}
Antisymmetrisation has unit weight. Substituting \eqref{eq:appSevenCoset}
and \eqref{eq:appSO8Embedding} into the maximal-supergravity potential \cite{deWit:2007kvg} given, in our conventions, by 
(2.27) of \cite{Guarino:2015qaa},
\begin{equation}
\label{eq:appMaximalPotential}
 V=\frac{g^2}{168}\,X_{MP}{}^R X_{NQ}{}^S\,\cM^{MN}
       \big(\cM^{PQ}\cM_{RS}+7\delta^P_S\delta^Q_R\big) \; ,
\end{equation}
gives the potential \eqref{eq:PotFromSuperPot} of the seven-chiral
sector. In the K\"ahler gauge of \eqref{eq:KahlerPot}, this potential
is reproduced by the holomorphic superpotential \eqref{SuperPot}.

\subsection{Eleven-dimensional conventions}
\label{app:D11Conventions}

The uplifted $D=11$ fields take the form \eqref{eq: upliftansatz} of the main text. With the four-dimensional vectors set to zero, the warp factor $\Delta$,
internal metric $\dd s_7^2$ and internal three-form potential $A_3$ follow
from (24)--(26) of \cite{Varela:2015ywx} as
\begin{align}
 \Delta^{-3}
 &=\tfrac{1}{84}\,X'_{MP}{}^R X'_{NQ}{}^S\,\cM^{MN}
       \big(\cM^{PQ}\cM_{RS}+7\delta^P_S\delta^Q_R\big) \; ,\notag\\*[3pt]
 \dd s_7^2
 &=\frac{\Delta^2}{12g^2}\,(t_I{}^J)_M{}^P(t_K{}^L)_N{}^Q\,
       \cM^{MN}\cM_{PQ}\,\mu_J\mu_L\,\dd\mu^I\dd\mu^K \; ,\label{eq:appGeneralUplift}\\*[3pt]
 A_3&=-\frac{\Delta^3}{72g^3}\,(t_I{}^J)_P{}^R X'_{MQ}{}^S\,\delta_{NT}\Omega^{TU}\Theta_U{}^K{}_L\,\cM^{MN}\cM^{PQ}\cM_{RS}\,\mu_J\,\dd\mu^I\wedge\dd\mu_K\wedge\dd\mu^L \; .\notag
\end{align}
Here $g$ is the SO(8) gauge coupling, and the embedding coordinates
$\mu^I$ obey \eqref{eq: S7inR8}, with $\mu_I=\delta_{IJ}\mu^J$.
The generators, scalar matrix and SO(8) embedding tensor are those of
appendix~\ref{app:D4Conventions}. In the electric and magnetic basis of
\eqref{eq:appSO8Embedding}, $\delta_{MN}$ is the scalar matrix at the
origin and $\Omega^{MN}$ is the invariant symplectic tensor, with
upper-right and lower-left blocks $\one_{28}$ and $-\one_{28}$ and
vanishing diagonal blocks. The primed tensor $X'_{MN}{}^P$ is obtained
from $X_{MN}{}^P$ in \eqref{eq:appSO8Embedding} by replacing
$\delta_{JL}$ in the electric embedding tensor with $\mu_J\mu_L$, namely,
\begin{equation}
\label{eq:appSO8EmbeddingPrimed}
 \Theta'_{[IJ]}{}^K{}_L
 =2\delta^K_{[I}\mu_{J]}\mu_L \; , \qquad
 \Theta'^{[IJ]K}{}_L=0 \; , \qquad
 X'_{MN}{}^P=\Theta'_M{}^I{}_J(t_I{}^J)_N{}^P \; .
\end{equation}
Specialising \eqref{eq:appGeneralUplift} to the coset representative
\eqref{eq:appSevenCoset} gives \eqref{eq:UpliftedExpr}, after extensive manipulation.

\section{Details on the seven-chiral $S^7$ uplift}
\label{sec:seven-chiral-uplift-blocks}


Here we make explicit all the building blocks, summarised in table~\ref{tab:uplift-building-blocks}, that enter the $D=11$ uplift discussed in section~\ref{sec:7CS11D} of the seven-chiral subsector of $D=4$ $\cN=8$ SO(8)-gauged supergravity. We first give the explicit expressions for convenience, and then comment on the Fano organisation underlying them.

\subsection{Explicit matrix blocks}
\label{sec:seven-chiral-explicit-blocks}

According to table~\ref{tab:uplift-building-blocks}, the quantities that depend only on the $S^7$ coordinates are the column vector $\mu$, the diagonal $8\times 8$ matrix $D$ and the seven $8\times 8$ matrices $Q_d$. The first two are given in terms of the embedding coordinates $\mu^I$ of $\mathbb{R}^8$, $I=1,\dots,8$, subject to~(\ref{eq: S7inR8}), by 
\begin{equation} \label{eq:sphere}
\mu=(\mu^1,\ldots,\mu^8)^{\T} \; , \qquad 
D=\diag(\mu^1,\ldots,\mu^8) \; ,
\end{equation}
so that $\mu^{\T}\mu=1$ and $\mu^{\T}\dd\mu=0$. As for $Q_d$, $d=1, \ldots, 7$, these are explicitly given in a convenient basis by
\begingroup
\small
\setlength{\arraycolsep}{2pt}
\begin{gather}
Q_1=\begin{pmatrix}
\mu^8&0&0&0&0&0&0&0\\
0&0&-\mu^2&0&0&0&0&0\\
0&\mu^3&0&0&0&0&0&0\\
0&0&0&0&0&0&0&-\mu^1\\
0&0&0&0&-\mu^4&0&0&0\\
0&0&0&0&0&0&-\mu^6&0\\
0&0&0&0&0&\mu^7&0&0\\
0&0&0&\mu^5&0&0&0&0
\end{pmatrix},\quad
Q_2=\begin{pmatrix}
0&\mu^8&0&0&0&0&0&0\\
0&0&\mu^1&0&0&0&0&0\\
-\mu^3&0&0&0&0&0&0&0\\
0&0&0&0&0&0&0&-\mu^2\\
0&0&0&0&0&\mu^4&0&0\\
0&0&0&0&0&0&-\mu^5&0\\
0&0&0&0&\mu^7&0&0&0\\
0&0&0&-\mu^6&0&0&0&0
\end{pmatrix},\notag\\[12pt]
Q_3=\begin{pmatrix}
0&0&\mu^8&0&0&0&0&0\\
0&-\mu^1&0&0&0&0&0&0\\
\mu^2&0&0&0&0&0&0&0\\
0&0&0&0&0&0&0&-\mu^3\\
0&0&0&0&0&0&-\mu^4&0\\
0&0&0&0&0&-\mu^5&0&0\\
0&0&0&0&\mu^6&0&0&0\\
0&0&0&\mu^7&0&0&0&0
\end{pmatrix},\quad
Q_4=\begin{pmatrix}
0&0&0&\mu^8&0&0&0&0\\
0&0&0&0&\mu^1&0&0&0\\
-\mu^5&0&0&0&0&0&0&0\\
0&0&0&0&0&0&0&-\mu^4\\
0&0&0&0&0&-\mu^2&0&0\\
0&0&0&0&0&0&\mu^3&0\\
0&0&-\mu^7&0&0&0&0&0\\
0&\mu^6&0&0&0&0&0&0
\end{pmatrix},\notag\\[12pt]
Q_5=\begin{pmatrix}
0&0&0&0&\mu^8&0&0&0\\
0&0&0&-\mu^1&0&0&0&0\\
\mu^4&0&0&0&0&0&0&0\\
0&0&0&0&0&0&0&-\mu^5\\
0&0&0&0&0&0&\mu^2&0\\
0&0&0&0&0&\mu^3&0&0\\
0&0&-\mu^6&0&0&0&0&0\\
0&-\mu^7&0&0&0&0&0&0
\end{pmatrix},\quad
Q_6=\begin{pmatrix}
0&0&0&0&0&\mu^8&0&0\\
0&0&0&0&0&0&\mu^1&0\\
-\mu^7&0&0&0&0&0&0&0\\
0&0&0&0&0&0&0&-\mu^6\\
0&0&0&\mu^2&0&0&0&0\\
0&0&0&0&-\mu^3&0&0&0\\
0&0&\mu^5&0&0&0&0&0\\
0&-\mu^4&0&0&0&0&0&0
\end{pmatrix},\notag\\[12pt]
Q_7=\begin{pmatrix}
0&0&0&0&0&0&\mu^8&0\\
0&0&0&0&0&-\mu^1&0&0\\
\mu^6&0&0&0&0&0&0&0\\
0&0&0&0&0&0&0&-\mu^7\\
0&0&0&0&-\mu^2&0&0&0\\
0&0&0&-\mu^3&0&0&0&0\\
0&0&\mu^4&0&0&0&0&0\\
0&\mu^5&0&0&0&0&0&0
\end{pmatrix}.
\label{eq:Qd}
\end{gather}
\endgroup

The quantities that depend only on the $D=4$ scalars are the seven $8\times 8$ symmetric matrices $T_d$, the $4\times 4$ matrices $U_d$, and the diagonal matrix $P$. These can all be written in terms of the seven real $D=4$ scalars $\varphi_i$ and seven pseudoscalars $\chi_i$ introduced in section~\ref{sec:7CS4D} through the combinations (\ref{eq:scalars}). In order to write $T_d$, we first define the following real $2 \times 2$ matrices
\begin{equation} \label{eq:ABC-blocks}
A_{ij}=\begin{pmatrix}a_ia_j&b_ib_j\\b_ib_j&c_ic_j\end{pmatrix},
\quad
B_{ij}=\begin{pmatrix}a_ib_j&b_ia_j\\b_ic_j&c_ib_j\end{pmatrix},
\quad
C_{ij}=\begin{pmatrix}a_ic_j&b_ib_j\\b_ib_j&c_ia_j\end{pmatrix} \; , 
\end{equation}
and the following collection of $8 \times 8$ matrices $\mathcal T_{ijk\ell}$, given as $4 \times 4$ arrays of the $2 \times 2$ blocks (\ref{eq:ABC-blocks}),
\begin{equation} \label{eq:template}
\mathcal T_{ijk\ell}=
\begin{pmatrix}
a_ia_jA_{k\ell}&b_ib_jA_{k\ell}&a_ib_jB_{k\ell}&b_ia_jB_{k\ell}\\
b_ib_jA_{k\ell}&c_ic_jA_{k\ell}&b_ic_jB_{k\ell}&c_ib_jB_{k\ell}\\
a_ib_jB_{k\ell}^{\T}&b_ic_jB_{k\ell}^{\T}&a_ic_jC_{k\ell}&b_ib_jC_{k\ell}\\
b_ia_jB_{k\ell}^{\T}&c_ib_jB_{k\ell}^{\T}&b_ib_jC_{k\ell}&c_ia_jC_{k\ell}
\end{pmatrix}.
\end{equation}
The matrices $T_d$, $d=1, \ldots, 7$, are finally given by
\begin{equation} \label{eq:Td}
\begin{gathered}
T_1=\mathcal T_{4567},\qquad
T_2=\mathcal T_{2367},\qquad
T_3=\mathcal T_{2345},\qquad
T_4=\mathcal T_{1357},\\
T_5=\mathcal T_{1346},\qquad
T_6=\mathcal T_{1256},\qquad
T_7=\mathcal T_{1247}.
\end{gathered}
\end{equation}
Similarly, introduce the universal $4\times 4$ matrix
\begin{equation} \label{eq:U-template}
\mathcal U_{ijk}=
\begin{pmatrix}
b_ib_jb_k&b_ia_ja_k&a_ib_ja_k&a_ia_jb_k\\
b_ic_jc_k&b_ib_jb_k&a_ic_jb_k&a_ib_jc_k\\
c_ib_jc_k&c_ia_jb_k&b_ib_jb_k&b_ia_jc_k\\
c_ic_jb_k&c_ib_ja_k&b_ic_ja_k&b_ib_jb_k
\end{pmatrix}.
\end{equation}
The matrices $U_d$, $d=1,\ldots,7$, are then given by
\begin{equation} \label{eq:Ud}
\begin{gathered}
U_1=\mathcal U_{123},\qquad
U_2=\mathcal U_{145},\qquad
U_3=\mathcal U_{167},\qquad
U_4=\mathcal U_{246},\\
U_5=\mathcal U_{257},\qquad
U_6=\mathcal U_{347},\qquad
U_7=\mathcal U_{356}.
\end{gathered}
\end{equation}
Finally, the matrix $P$ is 
\begin{equation}
P=\diag\left(b_1^2+\frac12,\ b_2^2+\frac12,\ b_3^2+\frac12,
\ b_4^2+\frac12,\ b_5^2+\frac12,\ b_6^2+\frac12,\ b_7^2+\frac12\right).
\label{eq:P}
\end{equation}

The last quantity from table~\ref{tab:uplift-building-blocks} that remains to be defined is the $7 \times 8$ matrix $S$. This is constant, independent of both the $S^7$ coordinates and the $D=4$ scalars, and is explicitly given by
\begin{equation}
S=\begin{pmatrix}
1&1&1&-1&-1&-1&-1&1\\
1&-1&-1&1&1&-1&-1&1\\
1&-1&-1&-1&-1&1&1&1\\
-1&1&-1&1&-1&1&-1&1\\
-1&1&-1&-1&1&-1&1&1\\
-1&-1&1&1&-1&-1&1&1\\
-1&-1&1&-1&1&1&-1&1
\end{pmatrix}.\label{eq:S}
\end{equation}

The quantities ${\cal H}$, $\alpha_d$ and $\beta_d$ that manifestly enter the uplift expressions (\ref{eq:UpliftedExpr}) can all be derived from the quantities defined in this appendix. Specifically, the $8 \times 8$ symmetric matrix ${\cal H}$ depends both on the $D=4$ scalars and the $S^7$ angles, and is obtained by substituting the above expressions for $Q_d$, $T_d$, $D$, $S$ and $P$ into (\ref{eq:H}). The four-component vectors $\alpha_d$ of one-forms also depend on the $D=4$ scalars and the $S^7$ angles, and can be obtained as follows. Introduce the eight-dimensional vector 
\begin{equation} \label{eq:qOneForm}
q=\HH \, \dd\mu \; , 
\end{equation}
with one-form components $q^I$, $I=1, \ldots, 8$, on $S^7$, and define the combinations
\begin{equation} \label{eq:alphaComb}
\alpha^{IJ} \equiv \mu^I q^J -\mu^J q^I \; .
\end{equation}
Then $\alpha_d$, $d=1, \ldots, 7$, are extracted from (\ref{eq:alphaComb}) by
\begin{equation}
\label{eq:alpha-columns}
\begin{gathered}
\alpha_1=\begin{pmatrix}\alpha^{18}\\\alpha^{23}\\\alpha^{45}\\\alpha^{67}\end{pmatrix},\quad
\alpha_2=\begin{pmatrix}\alpha^{28}\\\alpha^{31}\\\alpha^{64}\\\alpha^{57}\end{pmatrix},\quad
\alpha_3=\begin{pmatrix}\alpha^{38}\\\alpha^{12}\\\alpha^{47}\\\alpha^{56}\end{pmatrix},\quad
\alpha_4=\begin{pmatrix}\alpha^{48}\\\alpha^{51}\\\alpha^{26}\\\alpha^{73}\end{pmatrix},\\[8pt]
\alpha_5=\begin{pmatrix}\alpha^{58}\\\alpha^{14}\\\alpha^{72}\\\alpha^{63}\end{pmatrix},\quad
\alpha_6=\begin{pmatrix}\alpha^{68}\\\alpha^{71}\\\alpha^{42}\\\alpha^{35}\end{pmatrix},\quad
\alpha_7=\begin{pmatrix}\alpha^{78}\\\alpha^{16}\\\alpha^{25}\\\alpha^{34}\end{pmatrix}.
\end{gathered}
\end{equation}
Finally, $\beta_d$ depend only on the $S^7$ angles and not on the $D=4$ scalars. Define the two-forms on $S^7$
\begin{equation} \label{eq:betaComb}
\beta^{IJ} \equiv \d\mu^I \wedge \d\mu^J \; ,
\end{equation}
Then the vectors $\beta_d$ are obtained from (\ref{eq:betaComb}) using the same ordered pairs as $\alpha_d$ in (\ref{eq:alpha-columns}), {\it i.e.}, 
$\beta_1^{\T} = \big( \beta^{18} , \beta^{23} , \beta^{45} , \beta^{67} \big) $, etc.

\subsection{Fano mnemonic}
\label{sec:Fano-seven-chiral-uplift}

The Fano plane provides a mnemonic for the scalar dependence and coordinate-pair organisation of the building blocks above. Label its seven points by $1,\ldots,7$, and order its seven lines $L_d$ as in table~\ref{tab:Fano-uplift}. Each line contains three points, and its complement $L_d^c$ contains the remaining four. As already noted in section~\ref{sec:7CS4D}, these are the same seven triples that enter the cubic terms of the superpotential (\ref{SuperPot}), which can be written as
\begin{equation}
W=2g\left[\sum_{d=1}^{7}\prod_{i\in L_d}t_i+\prod_{i=1}^{7}t_i\right].
\label{eq:Fano-superpotential}
\end{equation}
The uplift uses both these lines and their complements.

\begin{table}[htbp]
\centering
\renewcommand{\arraystretch}{1.15}
\begin{tabular}{c c c c}
\toprule
$d$ & Scalar line $L_d$ & Complement $L_d^c$ & Ordered coordinate pairs \\
\midrule
1 & $123$ & $4567$ & $18,\ 23,\ 45,\ 67$ \\
2 & $145$ & $2367$ & $28,\ 31,\ 64,\ 57$ \\
3 & $167$ & $2345$ & $38,\ 12,\ 47,\ 56$ \\
4 & $246$ & $1357$ & $48,\ 51,\ 26,\ 73$ \\
5 & $257$ & $1346$ & $58,\ 14,\ 72,\ 63$ \\
6 & $347$ & $1256$ & $68,\ 71,\ 42,\ 35$ \\
7 & $356$ & $1247$ & $78,\ 16,\ 25,\ 34$ \\
\bottomrule
\end{tabular}
\caption{Fano organisation of the uplift blocks. The scalar labels in $U_d$ lie on $L_d$, while those in $T_d$ lie on $L_d^c$. The last column gives the ordered, oriented pairs used in $\alpha_d$ and $\beta_d$.}
\label{tab:Fano-uplift}
\end{table}

For each $d$, every entry of $U_d$ contains one factor $a_i$, $b_i$ or $c_i$ from each of the three labels $i\in L_d$. Similarly, every entry of $T_d$ contains one factor from each of the four labels in $L_d^c$. This explains the cubic and quartic scalar-factor structure of (\ref{eq:U-template}) and (\ref{eq:template}), respectively. The tensor-product origin of the latter can be made explicit by defining
\begin{equation}
m_i=\begin{pmatrix}a_i&b_i\\b_i&c_i\end{pmatrix},
\qquad \det m_i=a_ic_i-b_i^2=1,
\label{eq:Fano-scalar-matrices}
\end{equation}
in terms of the shorthand scalar combinations (\ref{eq:scalars}). For $L_d^c=\{i,j,k,\ell\}$, with $i<j<k<\ell$, the matrix $T_d=\mathcal T_{ijk\ell}$ is the principal submatrix of $m_i\otimes m_j\otimes m_k\otimes m_\ell$ selected by the even-parity binary strings, in the order
\begin{equation}
(0000,\ 0011,\ 1100,\ 1111,\ 0101,\ 0110,\ 1001,\ 1010).
\label{eq:Fano-even-parity-order}
\end{equation}
Here even parity means that an even number of entries are equal to $1$. This order reproduces the template (\ref{eq:template}) and is the row order used by the corresponding $Q_d$ in (\ref{eq:Qd}).

The same lines determine the signs in the constant matrix $S$, (\ref{eq:S}). Row $d$ has positive entries at the three points of $L_d$ and at the eighth coordinate, with negative entries on the complement:
\begin{equation}
S_{dI}=\begin{cases}
+1,& I\in L_d\ \textrm{or}\ I=8,\\
-1,& I\in L_d^c.
\end{cases}
\label{eq:Fano-S-signs}
\end{equation}
Together with $P$ and $D$, this gives the contribution $\tfrac14DS^{\T}PSD$ to ${\cal H}$ in (\ref{eq:H}).

The coordinate pairs have a related incidence description. Fix a point $d$ and take the pair $(d,8)$. There are three Fano lines through $d$; removing $d$ from each leaves three pairs, which complete the last column of table~\ref{tab:Fano-uplift}. With the displayed order and orientations, these select the four components of $\alpha_d$ and $\beta_d$ in (\ref{eq:alpha-columns}) and (\ref{eq:betaComb}). The same pairing pattern underlies the sparse coordinate matrices $Q_d$.

The index $d$ thus has two distinct roles: it enumerates the lines $L_d$ for the scalar blocks and the rows of $S$, whereas it labels a point in the coordinate-pair rule. The association between these roles is fixed by the table. For example, $L_2=\{1,4,5\}$ gives the scalar labels in $U_2$, the complement $\{2,3,6,7\}$ gives $T_2$, and $S_{2I}=(+1,-1,-1,+1,+1,-1,-1,+1)$. The coordinate pairs for point $2$, however, come from the three lines $123$, $246$ and $257$ through that point and are ordered as $(28,31,64,57)$ after including $(2,8)$.

The orientations in the table are part of the convention: for example, $\alpha^{31}=-\alpha^{13}$ and $\beta^{31}=-\beta^{13}$. Fano incidence determines the unordered pair grouping; the displayed orientations and the row and column orders of the explicit matrices fix the signs and ordering needed in the uplift.


\section{Metric and flux of the $\mathrm{U}(1)^2$ solution}
\label{app:U12MetricFlux}


We collect here the metric coefficients and internal three-form for the U$(1)^2$-invariant $D=11$ configuration, using the angular coordinates of section~\ref{sec:U12Sol11D}, the functions (\ref{eq:U12XY}), (\ref{eq:UpliftQuantities}), the one-forms (\ref{eq:U12omega}) and the warp factor (\ref{eq:U12CLPwarp}). These expressions hold throughout the two-scalar space parametrised by $\lambda_0,\lambda$ and give the $\cN=1$ U$(1)^2$-invariant solution upon substitution of the vacuum values (\ref{Field_IdsTwoScal}). At the $\cN=8$ SO(8)-invariant vacuum, $\lambda_0=\lambda=0$, the coefficients $B_i,U_i,V_i,M_{ij}$ and internal forms given below all vanish.

\subsection{Metric coefficients}

In order to define the remaining metric functions, it is convenient to introduce the following four functions $B_i$, $i=1, \ldots, 4$, of the $D=4$ scalar $\lambda$ and the $\mathbb{CP}^3$ angles $(\xi, \theta,\widetilde\theta,\phi,\widetilde\phi)$:
\begin{equation}
\begin{aligned}
B_1={}&\frac{\sinh(2\lambda)}{2\sqrt2}
 \cos\theta\sin\phi\cos^2\xi,\\[0.5ex]
B_2={}&\frac{\sinh\lambda}{\sqrt2}
 (X\cos\phi-Y\sin\phi)\cos^2\xi,\\[0.5ex]
B_3={}&\frac{\sinh(2\lambda)}{2\sqrt2}
 \cos\widetilde\theta\sin\widetilde\phi\sin^2\xi,\\[0.5ex]
B_4={}&\frac{\sinh\lambda}{\sqrt2}
 (\widetilde X\cos\widetilde\phi-\widetilde Y\sin\widetilde\phi)\sin^2\xi.
\end{aligned}
\label{eq:U12BcomponentsExplicit}
\end{equation}
With these definitions, the coefficients $U_i$, $V_i$, $i=1, \ldots, 4$, that enter the connections $\mathcal A$ in (\ref{eq:U12fluxConnection}) and $\widetilde{\mathcal A}$ in (\ref{eq:U12RelativeFibre}) are (note that these functions are unrelated to the matrices (\ref{eq:Ud}))
\begin{align}
U_1={}&\tfrac{\sinh\lambda}{\sqrt2}\Bigl[ \cosh\lambda(1+\tfrac12\sinh^2\lambda\cos^2\theta) \sin\phi\cos^2\xi +\cosh(\sqrt2\lambda_0) (\widetilde X\sin\phi+\widetilde Y\cos\phi)\sin^2\xi\Bigr],\nonumber\\[0.5ex]
U_2={}&\cos\theta\Bigl[ (1+\tfrac12\sinh^2\lambda)B_2 +\frac{\sinh\lambda}{\sqrt2}\cosh(\sqrt2\lambda_0) (\widetilde X\cos\phi-\widetilde Y\sin\phi)\sin^2\xi\Bigr],\nonumber\\[0.5ex]
U_3={}&\frac12\sinh^2\lambda\cos\theta\,B_3 -\sinh(\sqrt2\lambda_0)B_4,\nonumber\\[0.5ex]
U_4={}&\frac12\sinh^2\lambda\cos\theta\,B_4 +\sinh(\sqrt2\lambda_0)B_3 ,
\label{eq:U12UcomponentsExplicit}
\end{align}
and
\begin{align}
V_1={}&\frac12\sinh^2\lambda\cos\widetilde\theta\,B_1 +\sinh(\sqrt2\lambda_0)B_2,\nonumber\\[0.5ex]
V_2={}&\frac12\sinh^2\lambda\cos\widetilde\theta\,B_2 -\sinh(\sqrt2\lambda_0)B_1,\nonumber\\[0.5ex]
V_3={}&\tfrac{\sinh\lambda}{\sqrt2}\Bigl[ \cosh\lambda(1+\tfrac12\sinh^2\lambda\cos^2\widetilde\theta) \sin\widetilde\phi\sin^2\xi +\cosh(\sqrt2\lambda_0) (X\sin\widetilde\phi+Y\cos\widetilde\phi)\cos^2\xi\Bigr],\nonumber\\[0.5ex]
V_4={}&\cos\widetilde\theta\Bigl[ (1+\tfrac12\sinh^2\lambda)B_4 +\frac{\sinh\lambda}{\sqrt2}\cosh(\sqrt2\lambda_0) (X\cos\widetilde\phi-Y\sin\widetilde\phi)\cos^2\xi\Bigr].
\label{eq:U12VcomponentsExplicit}
\end{align}
Finally, the metric functions $M_{ij}$ that appear in the deformed metric \eqref{eq:U12sixmetric} on $\mathbb{CP}^3$ can be explicitly written in terms of $B_i$, $U_i$ and $V_i$ in (\ref{eq:U12BcomponentsExplicit})--(\ref{eq:U12VcomponentsExplicit}) as 
\begin{equation}
M_{ij}= M_{ji} = B_iB_j-
\frac{Q U_iU_j\cos^2\xi+P V_iV_j\sin^2\xi
 -R(U_iV_j+V_iU_j)\cos^2\xi\sin^2\xi}
 {PQ-R^2\cos^2\xi\sin^2\xi}.
\label{eq:U12Mcomponents}
\end{equation}
The quantities (\ref{eq:U12BcomponentsExplicit})--(\ref{eq:U12Mcomponents}) depend on the $D=4$ scalars $\lambda_0$, $\lambda$, and on the $\mathbb{CP}^3$ angles $(\xi, \theta,\widetilde\theta,\phi,\widetilde\phi)$. They are independent of the remaining $\mathbb{CP}^3$ angle $\widetilde\psi$ and of  the $S^7$ Hopf angle $\psi$.

\subsection{Internal three-form}

With the overall factor $g^{-3}\Delta^3$ displayed explicitly in
\eqref{eq:U12FluxDecompositionMain}, the component forms can be written
in the following order: $C_1$, then $C_2$, then $\widetilde C_2$, and
finally $C_3$. The expressions use the preceding component forms, together with the
functions and one-forms defined above and in section~\ref{sec:U12Sol11D}.

The one-form component is
\begingroup
\renewcommand{\baselinestretch}{1}
\small
\setlength{\jot}{2pt}
\thinmuskip=2mu
\medmuskip=3mu
\thickmuskip=4mu
\begin{align}
C_1={}&-\frac{\Delta^{-3}\cosh\lambda\,(Y+\widetilde Y)\cos\xi\sin\xi}
 {2\bigl(X\widetilde X+Y\widetilde Y\bigr)}\,\dd\xi
 +\frac{P\sin^2\xi}{4}(B_1\omega^1+B_2\omega^2)
 -\frac{Q\cos^2\xi}{4}(B_3\omega^3+B_4\omega^4)\nonumber\\[0pt]
&-\frac{\cosh\lambda\,(Y-\widetilde Y)\cos^2\xi\sin^2\xi}{4}
 (U_2\omega^1-U_1\omega^2+V_4\omega^3-V_3\omega^4)\nonumber\\[0pt]
&+\frac{\sinh^2\lambda\cos^2\xi\sin^2\xi}{8}\Bigl\lbrace
 \bigl[\cos\theta\cosh\lambda\,(Y-\widetilde Y)
       +\cos\widetilde\theta\sinh(\sqrt2\lambda_0)\bigr]
 (B_2\omega^1-B_1\omega^2)\nonumber\\[0pt]
&\qquad+\bigl[\cos\widetilde\theta\cosh\lambda\,(Y-\widetilde Y)
       +\cos\theta\sinh(\sqrt2\lambda_0)\bigr]
 (B_4\omega^3-B_3\omega^4)\nonumber\\[0pt]
&\qquad-\cos\theta(U_1\omega^1+U_2\omega^2)
 +\cos\widetilde\theta(V_3\omega^3+V_4\omega^4)\Bigr\rbrace.
\label{eq:U12FluxC1}
\end{align}
\endgroup
The two-form component $C_2$ is
\begingroup
\renewcommand{\baselinestretch}{1}
\small
\setlength{\jot}{2pt}
\thinmuskip=2mu
\medmuskip=3mu
\thickmuskip=4mu
\begin{align}
C_2={}&\frac{\Delta^{-3}\,\dd\xi}{2\bigl(X\widetilde X+Y\widetilde Y\bigr)}
 \wedge\bigl[-\tan\xi(B_2\omega^1-B_1\omega^2)
             +\cot\xi(B_4\omega^3-B_3\omega^4)\bigr]\nonumber\\[0pt]
&-\frac{(Q+R\cos^2\xi)U_i-(P+R\sin^2\xi)V_i}
 {PQ-R^2\cos^2\xi\sin^2\xi}\,\omega^i\wedge C_1\nonumber\\[0pt]
&+\frac14\Bigl\lbrace
 \Bigl[\sinh(\sqrt2\lambda_0)\cos^2\xi
       (\Theta\cos^2\xi+B_1^2+B_2^2)\nonumber\\[0pt]
&\qquad\quad+\sin^2\xi\bigl(B_2U_1-B_1U_2
       +\cosh\lambda\,(Y-\widetilde Y)\Theta\cos^2\xi\bigr)\Bigr]
 \omega^1\wedge\omega^2\nonumber\\[0pt]
&\qquad+\Bigl[\cos^2\xi\bigl(B_4V_3-B_3V_4
       -\cosh\lambda\,(Y-\widetilde Y)\widetilde\Theta\sin^2\xi\bigr)\nonumber\\[0pt]
&\qquad\quad-\sinh(\sqrt2\lambda_0)\sin^2\xi
       (\widetilde\Theta\sin^2\xi+B_3^2+B_4^2)\Bigr]
 \omega^3\wedge\omega^4\nonumber\\[0pt]
&\qquad+(B_2\omega^1-B_1\omega^2)\wedge
 \bigl[(U_4\cos^2\xi+V_4\sin^2\xi)\omega^3
       -(U_3\cos^2\xi+V_3\sin^2\xi)\omega^4\bigr]\nonumber\\[0pt]
&\qquad-\bigl[(U_2\cos^2\xi+V_2\sin^2\xi)\omega^1
       -(U_1\cos^2\xi+V_1\sin^2\xi)\omega^2\bigr]
 \wedge(B_4\omega^3-B_3\omega^4)\Bigr\rbrace.
\label{eq:U12FluxC2}
\end{align}
\endgroup
The other two-form component is
\begingroup
\renewcommand{\baselinestretch}{1}
\small
\setlength{\jot}{2pt}
\thinmuskip=2mu
\medmuskip=3mu
\thickmuskip=4mu
\begin{align}
\widetilde C_2={}&
 \frac{\Delta^{-3}\cot\xi\,\dd\xi}{4\bigl(X\widetilde X+Y\widetilde Y\bigr)}
 \wedge(B_4\omega^3-B_3\omega^4)
 -\frac{\cos^2\xi(P+R\sin^2\xi)}{2f}\,C_2\nonumber\\[0pt]
&+\frac{PV_i-R\cos^2\xi\,U_i}{2\bigl(PQ-R^2\cos^2\xi\sin^2\xi\bigr)}
 \omega^i\wedge C_1\nonumber\\[0pt]
&+\frac{\cos^2\xi}{8}\Bigl\lbrace
 \sinh(\sqrt2\lambda_0)(\Theta\cos^2\xi+B_1^2+B_2^2)
 \omega^1\wedge\omega^2\nonumber\\[0pt]
&\qquad+\bigl[B_4V_3-B_3V_4
       -\cosh\lambda\,(Y-\widetilde Y)\widetilde\Theta\sin^2\xi\bigr]
 \omega^3\wedge\omega^4\nonumber\\[0pt]
&\qquad+(B_2\omega^1-B_1\omega^2)\wedge(U_4\omega^3-U_3\omega^4)
 -(U_2\omega^1-U_1\omega^2)\wedge(B_4\omega^3-B_3\omega^4)
 \Bigr\rbrace.
\label{eq:U12FluxCtilde2}
\end{align}
\endgroup
Finally, the three-form component is
\begingroup
\renewcommand{\baselinestretch}{1}
\small
\setlength{\jot}{2pt}
\thinmuskip=2mu
\medmuskip=3mu
\thickmuskip=4mu
\begin{align}
C_3={}&\frac18
 \bigl(\Theta\cos^2\xi\,\omega^1\wedge\omega^2
       +\widetilde\Theta\sin^2\xi\,\omega^3\wedge\omega^4\bigr)
 \wedge(B_2\omega^1-B_1\omega^2+B_4\omega^3-B_3\omega^4)\nonumber\\[0pt]
&-\frac{U_i\cos^2\xi+V_i\sin^2\xi}{2f}\,\omega^i\wedge C_2
 -\frac{(Q+R\cos^2\xi)U_i-(P+R\sin^2\xi)V_i}
 {PQ-R^2\cos^2\xi\sin^2\xi}\,\omega^i\wedge\widetilde C_2\nonumber\\[0pt]
&+\frac{U_iV_j}{2\bigl(PQ-R^2\cos^2\xi\sin^2\xi\bigr)}
 \omega^i\wedge\omega^j\wedge C_1.
\label{eq:U12FluxC3}
\end{align}
\endgroup

\section{Metric, flux and limits of the $\mathrm{SO}(3)$ solution}
\label{app:SO3Details}

We give here the coefficients of the internal metric
\eqref{eq:introBerger} and three-form \eqref{eq:SO3FluxDecompositionMain}
for the SO(3)-invariant sector, using the scalar
combinations \eqref{eq:scalars}, the coordinates
\eqref{eq:SO3Bergerslice}, the
one-forms \eqref{eq:SO3BergerMC}, the functions \eqref{eq:SO3XYZ}
and the warp factor \eqref{eq:SO3Bergerwarp}. The expressions apply to arbitrary values
of the three complex scalars $t_1,t_4,t_5$ and give the new
$\cN=1$ solution at their SO(3)-invariant values in
\eqref{Field_Identifications}. We also recover the SO(8), G$_2$ and CPW solutions. 

\subsection{Metric coefficients} \label{sec:MetCoefsSO3}

The six independent components of the dimensionless orbit metric $\gamma_{ij}=\gamma_{ji}$
in \eqref{eq:introBerger} are
\begingroup
\renewcommand{\baselinestretch}{1}
\small
\setlength{\jot}{2pt}
\thinmuskip=2mu
\medmuskip=3mu
\thickmuskip=4mu
\allowdisplaybreaks[4]
\begin{align}
&\gamma_{12}=\frac{a_1b_1(b_4-b_5)}{2} \sin\xi\cos^3\xi\sin2\zeta\sin\zeta \big(a_5\sin\psi\cos\tau-c_5\cos\psi\sin\tau\big),\nonumber\\[3pt]
&\gamma_{13}= -\frac{a_1b_1(b_4-b_5)}{2} \sin\xi\cos^3\xi\sin2\zeta\cos\zeta \big(a_5\sin\psi\sin\tau+c_5\cos\psi\cos\tau\big),\nonumber\\[3pt]
&\gamma_{22}= a_1\cos^2\xi\cos^2\zeta \Big[c_4X +a_5Y +\big[(c_5-c_4)X +(a_4-a_5)Y\big]\sin^2\tau -b_5(b_4-b_5)Z\sin2\tau\Big],\nonumber\\[3pt]
&\gamma_{23}=\frac{a_1\cos^2\xi\sin2\zeta}{4} \Big[\big\{(c_5-c_4)X +(a_4-a_5)Y\big\}\sin2\tau -2b_5(b_4-b_5)Z\cos2\tau\Big],\nonumber\\[3pt]
&\gamma_{33}= -\gamma_{22}\tan^2\zeta +a_1\cos^2\xi\sin^2\zeta \big[(c_4+c_5)X +(a_4+a_5)Y\big],\nonumber\\[3pt]
&\gamma_{11}=\gamma_{22} +\gamma_{33} -\frac{\cos^4\xi\sin^22\zeta}{4} \Big[(b_1^2-b_5^2)(a_4c_5+a_5c_4) +2a_5c_5(b_1^2-b_4b_5)\Big].
\label{eq:SO3Bergerorbitblocks}
\end{align}
\endgroup

Using the inverse $\gamma^{ij}$ of the metric (\ref{eq:SO3Bergerorbitblocks}), the coefficients of the connection $\mathcal{A}^i =\mathcal{A}^i{}_a \, \dd q^a$, with $i=1,2,3$ and $q^a=(\xi,\zeta,\psi,\tau)$, are as follows: 
\begingroup
\renewcommand{\baselinestretch}{1}
\small
\setlength{\jot}{2pt}
\thinmuskip=2mu
\medmuskip=3mu
\thickmuskip=4mu
\allowdisplaybreaks[4]
\begin{align}
&\mathcal A^i{}_{\xi}=\left(\tan\xi+\frac{c_1a_5}{a_1c_5}\cot\xi\right) \left(\gamma^{i2}\gamma_{13}-\gamma^{i3}\gamma_{12}\right) +\frac{\gamma^{i1}\sin\xi\cos^3\xi\sin2\zeta}{4} \Big\{\big[(b_1^2-b_5^2)(c_4c_5-a_4a_5)\nonumber\\*
&\quad +(b_1^2-b_4b_5)(c_5^2-a_5^2)\big]\sin2\psi -(b_1^2-b_5^2)(a_4c_5-a_5c_4)\cos2\zeta\sin2\tau\Big\}\nonumber\\*
&\quad +\frac{b_1c_1(b_4-b_5)(a_5^2-c_5^2)\cos^4\xi\sin2\zeta\sin\psi}{2c_5} \left(\gamma^{i2}\cos\zeta\sin\tau+\gamma^{i3}\sin\zeta\cos\tau\right),\nonumber\\[3pt]
&\mathcal A^i{}_{\zeta}=\gamma^{i1}\Bigg[ \frac{2\gamma_{23}}{\sin2\zeta} -\frac{\cos^4\xi\sin^22\zeta}{4} (b_1^2-b_5^2)(a_4c_5-a_5c_4)\sin2\tau\Bigg] -\gamma^{i2}\gamma_{13}\cot\zeta -\gamma^{i3}\gamma_{12}\tan\zeta,\nonumber\\[3pt]
&\mathcal A^i{}_{\tau}=\gamma^{i1}\Bigg[ \frac{\cos2\zeta(\gamma_{22}-\gamma_{33})-\gamma_{11}}{\sin2\zeta} +\frac{\cos^4\xi\sin2\zeta\cos2\zeta}{4} (b_1^2-b_5^2)(a_4c_5-a_5c_4)\cos2\tau\Bigg]\nonumber\\*
&\quad -\gamma^{i2}\gamma_{12}\cot\zeta -\gamma^{i3}\gamma_{13}\tan\zeta,\nonumber\\[3pt]
&\mathcal A^i{}_{\psi}= -\frac{c_1}{a_1}\mathcal A^i{}_{\tau} -\frac{\gamma^{i1}\cos^2\xi\sin2\zeta}{2a_1} \Big[a_5c_5(c_4X+a_4Y) +(1+b_4b_5)(c_5X+a_5Y)\Big].
\label{eq:SO3Bergerblocks}
\end{align}
\endgroup

Finally, the ten independent components of $h_{ab} = h_{ba}$
in \eqref{eq:introBerger} are, with repeated indices $i,j=1,2,3$ summed,  
\begingroup
\renewcommand{\baselinestretch}{1}
\small
\setlength{\jot}{2pt}
\thinmuskip=2mu
\medmuskip=3mu
\thickmuskip=4mu
\allowdisplaybreaks[4]
\begin{align}
&h_{\xi\xi}= (c_4X+a_5Y)\Big[\frac{a_1\sin^2\xi}{2}(1+\cos2\zeta\cos2\tau) +\frac{c_1c_5}{a_5}\cos^2\xi\sin^2\psi\Big]\nonumber\\*
&\quad +(c_5X+a_4Y)\Big[\frac{a_1\sin^2\xi}{2}(1-\cos2\zeta\cos2\tau) +\frac{c_1a_5}{c_5}\cos^2\xi\cos^2\psi\Big]\nonumber\\
&\quad +b_5(b_4-b_5)Z\big(c_1\cos^2\xi\sin2\psi -a_1\sin^2\xi\cos2\zeta\sin2\tau\big)\nonumber\\
&\quad +\frac{\sin^2\xi\cos^2\xi}{4}\Bigg\{ (b_4-b_5)(b_4+b_5+2b_1^2b_5)(\cos2\psi+\cos2\zeta\cos2\tau)^2\nonumber\\
&\quad -(b_1^2-b_5^2)\Big[\frac{a_4}{c_5}(a_5\sin2\psi+c_5\cos2\zeta\sin2\tau)^2 +\frac{c_4}{a_5}(c_5\sin2\psi+a_5\cos2\zeta\sin2\tau)^2\Big]\nonumber\\*
&\quad +2(b_1^2-b_4b_5)(a_5\sin2\psi+c_5\cos2\zeta\sin2\tau) (c_5\sin2\psi+a_5\cos2\zeta\sin2\tau)\Bigg\} -\gamma_{ij}\mathcal A^i{}_{\xi}\mathcal A^j{}_{\xi},\nonumber\\[3pt]
&h_{\xi\zeta}= \frac{\sin2\xi\sin2\zeta}{8}\Bigg\{ 2a_1\Big[\big((c_4-c_5)X+(a_5-a_4)Y\big)\cos2\tau -2b_5(b_4-b_5)Z\sin2\tau\Big]\nonumber\\*
&\quad +\cos^2\xi\Big[ (b_4-b_5)(b_4+b_5+2b_1^2b_5)\cos2\tau (\cos2\psi+\cos2\zeta\cos2\tau)\nonumber\\
&\quad -\sin2\tau\Big[ \big((b_1^2-b_5^2)a_4-(b_1^2-b_4b_5)a_5\big) (a_5\sin2\psi+c_5\cos2\zeta\sin2\tau)\nonumber\\*
&\quad +\big((b_1^2-b_5^2)c_4-(b_1^2-b_4b_5)c_5\big) (c_5\sin2\psi+a_5\cos2\zeta\sin2\tau)\Big]\Big]\Bigg\} -\gamma_{ij}\mathcal A^i{}_{\xi}\mathcal A^j{}_{\zeta},\nonumber\\[3pt]
&h_{\xi\psi}= \frac{\sin2\xi}{4}\Bigg\{ c_1\sin2\psi\Big[\frac{c_5}{a_5}(c_4X+a_5Y) -\frac{a_5}{c_5}(c_5X+a_4Y)\Big] +2c_1b_5(b_4-b_5)Z\cos2\psi\nonumber\\*
&\quad +\sin^2\xi\Big[ \Big(\frac{a_4a_5}{c_5}(b_1^2-b_5^2)\sin^2\psi+c_5(b_1^2-b_4b_5)\cos^2\psi\Big) (a_5\sin2\psi+c_5\cos2\zeta\sin2\tau)\nonumber\\
&\quad -\Big(\frac{c_4c_5}{a_5}(b_1^2-b_5^2)\cos^2\psi+a_5(b_1^2-b_4b_5)\sin^2\psi\Big) (c_5\sin2\psi+a_5\cos2\zeta\sin2\tau)\nonumber\\*
&\quad -\frac{(b_4-b_5)(b_4+b_5+2b_1^2b_5)}{2}\sin2\psi (\cos2\psi+\cos2\zeta\cos2\tau)\Big]\Bigg\} -\gamma_{ij}\mathcal A^i{}_{\xi}\mathcal A^j{}_{\psi},\nonumber\\[3pt]
&h_{\xi\tau}= -\frac{a_1}{c_1}h_{\xi\psi} +\frac{\sin2\xi}{4c_1}\Big[ \big(c_4c_5X+(1+b_4b_5)Y\big) (c_5\sin2\psi+a_5\cos2\zeta\sin2\tau)\nonumber\\*
&\quad -\big(a_4a_5Y+(1+b_4b_5)X\big) (a_5\sin2\psi+c_5\cos2\zeta\sin2\tau) -(b_4-b_5)^2Z(\cos2\psi+\cos2\zeta\cos2\tau)\Big]\nonumber\\*
&\quad +\frac{\cos^2\xi\sin2\zeta}{2c_1}\mathcal A^1{}_{\xi} \Big[a_5c_5(c_4X+a_4Y) +(1+b_4b_5)(c_5X+a_5Y)\Big],\nonumber\\[3pt]
&h_{\zeta\zeta}=\gamma_{22}\tan^2\zeta +\gamma_{33}\cot^2\zeta -\gamma_{ij}\mathcal A^i{}_{\zeta}\mathcal A^j{}_{\zeta} +\frac{\cos^4\xi\sin^22\zeta}{4}\Big\{ (b_4-b_5)(b_4+b_5+2b_1^2b_5)\cos^22\tau\nonumber\\*
&\quad +\big[2a_5c_5(b_1^2-b_4b_5) -(b_1^2-b_5^2)(a_4c_5+a_5c_4)\big]\sin^22\tau\Big\},\nonumber\\[3pt]
&h_{\zeta\tau}= -2\gamma_{23} -\gamma_{ij}\mathcal A^i{}_{\zeta}\mathcal A^j{}_{\tau} +\frac{\cos^4\xi\sin2\zeta\sin2\tau}{4}\Big\{ (b_1^2-b_5^2)\big[a_4c_5 -a_5c_4\nonumber\\*
&\quad -(a_4-a_5)(c_4-c_5)\cos2\zeta\cos2\tau\big] +(b_4-b_5)\big[a_1c_1(b_4+3b_5) -b_5^2(b_4-b_5)\big]\cos2\zeta\cos2\tau\Big\},\nonumber\\[3pt]
&h_{\zeta\psi}= -\frac{c_1}{a_1}h_{\zeta\tau} +\frac{\cos^2\xi\sin2\zeta}{2a_1}\Bigg\{ \big[c_5(a_5c_4-1-b_4b_5)X -a_5(a_4c_5-1-b_4b_5)Y\big]\sin2\tau\nonumber\\*
&\quad -(b_4-b_5)^2Z\cos2\tau +\mathcal A^1{}_{\zeta}\Big[a_5c_5(c_4X+a_4Y) +(1+b_4b_5)(c_5X+a_5Y)\Big]\Bigg\},\nonumber\\[3pt]
&h_{\tau\tau}= -\gamma_{11} -\frac{2\gamma_{1i}\mathcal A^i{}_{\tau}}{\sin2\zeta} -\gamma_{ij}\mathcal A^i{}_{\tau}\mathcal A^j{}_{\tau} +\frac{\cos^4\xi\cos^22\zeta}{4}\Big\{ 4a_5c_5(b_1^2-b_4b_5)\cos^22\tau\nonumber\\*
&\quad +\big[(b_1^2-b_5^2)(a_4c_5+a_5c_4+2(1+b_4b_5)) +(b_4-b_5)^2\big]\sin^22\tau\Big\},\nonumber\\[3pt]
&h_{\psi\tau}= -\frac{c_1}{a_1}h_{\tau\tau} +\frac{\cos^2\xi}{2a_1}\Big\{ (1+\mathcal A^1{}_{\tau}\sin2\zeta) \big[a_5c_5(c_4X+a_4Y) +(1+b_4b_5)(c_5X+a_5Y)\big]\nonumber\\*
&\quad -\cos2\zeta\Big[ \big[c_5(a_5c_4-1-b_4b_5)X -a_5(a_4c_5-1-b_4b_5)Y\big]\cos2\tau +(b_4-b_5)^2Z\sin2\tau\Big]\Big\},\nonumber\\[3pt]
&h_{\psi\psi}= -\frac{2c_1}{a_1}h_{\psi\tau} -\frac{c_1^2}{a_1^2}h_{\tau\tau} +\frac{1}{a_1^2}\Big[c_4c_5X^2 +a_4a_5Y^2 +2(1+b_4b_5)XY +(b_4-b_5)^2Z^2\Big]\nonumber\\*
&\quad -\frac{\gamma^{11}\cos^4\xi\sin^22\zeta}{4a_1^2} \Big[a_5c_5(c_4X+a_4Y) +(1+b_4b_5)(c_5X+a_5Y)\Big]^2.
\label{eq:SO3Bergerbase}
\end{align}
\endgroup

\subsection{Internal three-form}
\label{sec:SO3InternalThreeForm}

With the overall factor $g^{-3}\Delta^3$ extracted in
\eqref{eq:SO3FluxDecompositionMain}, we specify the dimensionless function $C_0$ and the forms
\begin{equation}
C_3=\frac1{3!}C_{3abc}\,\dd q^a\wedge\dd q^b\wedge\dd q^c,
\qquad
C_{2i}=\frac12C_{2iab}\,\dd q^a\wedge\dd q^b,
\qquad
C_1{}^i=C_1{}^i{}_a\,\dd q^a
\label{eq:SO3FluxBaseComponents}
\end{equation}
that enter the decomposition
\eqref{eq:SO3FluxDecompositionMain} of the internal $D=11$ three-form. We again use the functions
\eqref{eq:SO3XYZ} and the connection ${\cal A}^i$ defined in section~\ref{sec:MetCoefsSO3}.
The warp factor \eqref{eq:SO3Bergerwarp} enters only through the overall factor in \eqref{eq:SO3FluxDecompositionMain}. Repeated transverse indices $a,b,c=1,\ldots,4$ and orbit indices
$i,j,k=1,2,3$ are summed, as usual. We determine the forms successively, starting with
\begin{equation}
C_0=-\frac{a_1b_1}{4}\cos^4\xi\sin^22\zeta\Big[a_5c_5(c_4X+a_4Y)+(1+b_4b_5)(c_5X+a_5Y)\Big].
\label{eq:SO3FluxC0}
\end{equation}
Let us now specify $C_1{}^i$. For $C_1{}^1$ we obtain
\begingroup
\renewcommand{\baselinestretch}{1}
\small
\setlength{\jot}{2pt}
\thinmuskip=2mu
\medmuskip=3mu
\thickmuskip=4mu
\allowdisplaybreaks[4]
\begin{align}
&C_1{}^1+C_0\Bigl(\mathcal A^1+\frac{\dd\tau}{\sin2\zeta}\Bigr)
=\frac{a_1b_1\cos^2\xi\sin2\zeta}{8}\Bigg\{
4c_5\sin\xi\cos\psi\big[c_4c_5X+(1+b_4b_5)Y\big]\,\dd(\sin\xi\sin\psi)\nonumber\\*
&-4a_5\sin\xi\sin\psi\big[(1+b_4b_5)X+a_4a_5Y\big]\,\dd(\sin\xi\cos\psi)\nonumber\\
&+\big[c_5(1+b_4b_5-a_5c_4)X+a_5(a_4c_5-1-b_4b_5)Y\big]\,\dd(\cos^2\xi\cos2\zeta\sin2\tau)\nonumber\\
&-(b_4-b_5)^2Z\,\dd\big(\sin^2\xi\cos2\psi-\cos^2\xi\cos2\zeta\cos2\tau\big)\Bigg\}.
\label{eq:SO3FluxC11}
\end{align}
\endgroup
For $C_1{}^2$, we obtain 
\begingroup
\renewcommand{\baselinestretch}{1}
\small
\setlength{\jot}{2pt}
\thinmuskip=2mu
\medmuskip=3mu
\thickmuskip=4mu
\allowdisplaybreaks[4]
\begin{align}
&(C_1{}^2+C_0\mathcal A^2) ={}\cos\xi\sin\zeta \Bigg\{ \frac{\cos^2\xi}{2} \Big[\cos\tau\Big\{ X\bigl[2a_1a_5b_4c_1c_5\cos2\zeta\cos2\tau\nonumber\\
&-(b_4-b_5)\bigl(a_1c_1-\tfrac12b_5(b_4-b_5)\bigr)(1+\cos2\zeta\cos2\tau)\bigr] +a_4a_5Y\bigl[2a_1b_5c_1\cos2\zeta\cos2\tau\nonumber\\
&-(b_4-b_5)(1-\cos2\zeta\cos2\tau)\bigr] +(b_4-b_5)Z\cos2\zeta\sin2\tau\bigl[a_4(b_1^2-b_5^2)\nonumber\\
&+\tfrac12a_5b_5(b_4-b_5)\bigr] +\tfrac14a_5c_1(b_4-b_5)\bigl[2b_1^2 -b_5(b_4+b_5)\bigr]\cos^2\xi\sin^22\zeta\Big\}\nonumber\\
&+\sin\tau\Big\{ -b_5X\cos2\zeta\sin2\tau\bigl[a_1c_1(1+b_4b_5+a_5c_4) +\tfrac12(b_4-b_5)^2\bigr]\nonumber\\
&-a_5Y\cos2\zeta\sin2\tau\bigl[a_1a_4b_5c_1 +a_5(b_5+b_1^2b_4)\bigr] -a_5(b_4-b_5)Z\bigl[(b_1^2-b_5^2)(1+\cos2\zeta\cos2\tau)\nonumber\\
&+\tfrac12b_5(b_4-b_5)(1-\cos2\zeta\cos2\tau)\bigr]\Big\}\Big]\dd(\sin\xi\cos\psi) +\frac{\cos^2\xi}{2} \Big[\cos\tau\Big\{\nonumber\\
&-c_5X\cos2\zeta\sin2\tau\bigl[c_5(b_5+b_1^2b_4) +a_1b_5c_1c_4\bigr] -b_5Y\cos2\zeta\sin2\tau\bigl[a_1c_1(1+b_4b_5+a_4c_5)\nonumber\\
&+\tfrac12(b_4-b_5)^2\bigr] -c_5(b_4-b_5)Z\bigl[(b_1^2-b_5^2)(1-\cos2\zeta\cos2\tau) +\tfrac12b_5(b_4-b_5)(1+\cos2\zeta\cos2\tau)\bigr]\Big\}\nonumber\\
&+\sin\tau\Big\{ -c_4c_5X\bigl[2a_1b_5c_1\cos2\zeta\cos2\tau +(b_4-b_5)(1+\cos2\zeta\cos2\tau)\bigr]\nonumber\\
&-Y\bigl[2a_1a_5b_4c_1c_5\cos2\zeta\cos2\tau +(b_4-b_5)\bigl(a_1c_1-\tfrac12b_5(b_4-b_5)\bigr)(1-\cos2\zeta\cos2\tau)\bigr]\nonumber\\
&+(b_4-b_5)Z\cos2\zeta\sin2\tau\bigl[c_4(b_1^2-b_5^2) +\tfrac12c_5b_5(b_4-b_5)\bigr] +\tfrac14c_1c_5(b_4-b_5)\bigl[2b_1^2\nonumber\\
&-b_5(b_4+b_5)\bigr]\cos^2\xi\sin^22\zeta\Big\}\Big]\dd(\sin\xi\sin\psi) +\frac{a_1\sin\xi}{4c_1} \Big[\cos\tau\Big\{ a_1b_4c_1c_5^2X\cos\psi\nonumber\\
&+\bigl[a_4c_5(b_4+b_1^2b_5) -\tfrac12(b_4-b_5)\{2b_1^2 -b_5(b_4+b_5)\}\bigr]Y\cos\psi +\tfrac12(b_4-b_5)\bigl[2a_4(b_1^2-b_5^2) +a_5\{2b_1^2\nonumber\\
&-b_5(b_4+b_5)\}\bigr]Z\sin\psi\Big\} +\sin\tau\Big\{ -b_5\bigl[a_1c_1(1+b_4b_5) +\tfrac12(b_4-b_5)^2\bigr]X\sin\psi -a_1a_4a_5b_5c_1Y\sin\psi\nonumber\\
&+\tfrac12b_5c_5(b_4-b_5)^2Z\cos\psi\Big\}\Big]\dd\bigl[\cos^2\xi(1-\cos2\zeta\cos2\tau)\bigr] +\frac{a_1\sin\xi}{2c_1} \Big[\cos\tau\Big\{ \bigl[a_5c_5(b_5+b_1^2b_4)X\nonumber\\
&+a_1a_4a_5b_5c_1Y\bigr]\sin\psi -c_5(b_4-b_5)(b_1^2-b_5^2)Z\cos\psi\Big\} +\sin\tau\Big\{ \bigl[a_1b_5c_1c_4c_5X\nonumber\\
&+a_5c_5(b_5+b_1^2b_4)Y\bigr]\cos\psi -a_5(b_4-b_5)(b_1^2-b_5^2)Z\sin\psi\Big\}\Big]\dd(\cos^2\xi\cos2\zeta\sin2\tau) +\frac{a_1\sin\xi}{4c_1} \Big[\cos\tau\Big\{\nonumber\\
&-a_1b_5c_1c_4c_5X\cos\psi -b_5\bigl[a_1c_1(1+b_4b_5) +\tfrac12(b_4-b_5)^2\bigr]Y\cos\psi +\tfrac12a_5b_5(b_4-b_5)^2Z\sin\psi\Big\}\nonumber\\
&+\sin\tau\Big\{ \bigl[a_5c_4(b_4+b_1^2b_5) -\tfrac12(b_4-b_5)\{2b_1^2 -b_5(b_4+b_5)\}\bigr]X\sin\psi +a_1a_5^2b_4c_1Y\sin\psi\nonumber\\
&+\tfrac12(b_4-b_5)\bigl[2c_4(b_1^2-b_5^2) +c_5\{2b_1^2 -b_5(b_4+b_5)\}\bigr]Z\cos\psi\Big\}\Big]\dd\bigl[\cos^2\xi(1+\cos2\zeta\cos2\tau)\bigr] \Bigg\}.
\label{eq:SO3FluxC12}
\end{align}
\endgroup
Finally, the expression for $C_1{}^3+C_0\mathcal A^3$ follows from the
right-hand side of \eqref{eq:SO3FluxC12} by the replacements
\begin{equation}
\zeta\mapsto\frac\pi2-\zeta,
\qquad \tau\mapsto\tau+\frac\pi2,
\qquad \dd\zeta\mapsto-\dd\zeta,
\label{eq:SO3FluxExchange}
\end{equation}
with $X,Y,Z$ unchanged.

Moving on to the two-forms $C_{2i}$, for $C_{21}$ we get
\begingroup
\renewcommand{\baselinestretch}{1}
\small
\setlength{\jot}{2pt}
\thinmuskip=2mu
\medmuskip=3mu
\thickmuskip=4mu
\allowdisplaybreaks[4]
\begin{align}
&C_{21} +\epsilon_{1jk}\mathcal A^j \wedge C_1{}^k +\frac12C_0\epsilon_{1jk}\mathcal A^j \wedge\mathcal A^k =2\bigl(-\tan\xi\,\dd\xi+\cot2\zeta\,\dd\zeta\bigr) \wedge(C_1{}^1+C_0\mathcal A^1)\nonumber\\
&+\frac{2\tan\xi}{\sin2\zeta}C_0\,\dd\xi \wedge\Bigl(\frac{c_1}{a_1}\dd\psi-\dd\tau\Bigr) +\frac{b_1(b_4-b_5)^2\cos^2\xi\sin2\zeta}{8} \Bigl[a_5^2\sin\xi\sin\psi\,\dd(\sin\xi\cos\psi)\nonumber\\
&+c_5^2\sin\xi\cos\psi\,\dd(\sin\xi\sin\psi) -\frac{a_5c_5}{2}\,\dd(\cos^2\xi\cos2\zeta\sin2\tau)\Bigr]\nonumber\\
&\wedge\dd\bigl(\sin^2\xi\cos2\psi-\cos^2\xi\cos2\zeta\cos2\tau\bigr).
\label{eq:SO3FluxC21}
\end{align}
\endgroup
For $C_{22}$, we find 
\begingroup
\renewcommand{\baselinestretch}{1}
\small
\setlength{\jot}{2pt}
\thinmuskip=2mu
\medmuskip=3mu
\thickmuskip=4mu
\allowdisplaybreaks[4]
\begin{align}
&C_{22}+\epsilon_{2jk}\mathcal A^j\wedge C_1{}^k+\frac12C_0\epsilon_{2jk}\mathcal A^j\wedge\mathcal A^k ={}\frac{\cos^3\xi\cos\zeta}{2}\Bigg\{ -\sin\tau\Big[\bigl[X a_{1} a_{5} b_{4} c_{1} c_{5}\nonumber\\
&+ Y a_{4} a_{5} \bigl(b_4+b_1^2b_5\bigr)\bigr]\,\dd(\sin\xi\cos\psi) +Z c_{5} \bigl(b_{1}^{2} - b_{5}^{2}\bigr) \bigl(b_{4} - b_{5}\bigr)\,\dd(\sin\xi\sin\psi)\Big] \wedge\Big[ -\sin2\tau\,\dd\zeta\nonumber\\
&+\sin2\zeta\,\dd\tau\Big] -\sin\tau\Big[Z a_{4} \bigl(b_{1}^{2} - b_{5}^{2}\bigr) \bigl(b_{4} - b_{5}\bigr)\,\dd(\sin\xi\cos\psi) +\bigl[ - X c_{5}^{2} \bigl(b_5+b_1^2b_4\bigr)\nonumber\\
&- Y a_{1} a_{4} b_{5} c_{1} c_{5}\bigr]\,\dd(\sin\xi\sin\psi)\Big] \wedge\Big[ -\tan\xi\sin2\zeta\,\dd\xi +(\cos2\zeta+\cos2\tau)\,\dd\zeta\Big]\nonumber\\
&+\Big[\frac{Z a_{5} b_{5} \bigl(b_{4} - b_{5}\bigr)^{2}}{2}\,\dd(\sin\xi\cos\psi) -b_5\bigl[a_1c_1c_4c_5X\nonumber\\
&+\bigl(a_1c_1(1+b_4b_5)+\tfrac12(b_4-b_5)^2\bigr)Y\bigr]\,\dd(\sin\xi\sin\psi)\Big] \wedge\Big[ -\sin\tau\tan\xi\sin2\zeta\,\dd\xi\nonumber\\
&+(\sin\tau\cos2\zeta-\sin3\tau)\,\dd\zeta +\cos\tau\sin2\zeta\,\dd\tau\Big] +\Big[\bigl[ -Xb_5\bigl[a_1c_1(1+b_4b_5) +\tfrac12(b_4-b_5)^2\bigr]\nonumber\\
&- Y a_{1} a_{4} a_{5} b_{5} c_{1}\bigr]\,\dd(\sin\xi\cos\psi) +\frac{Z b_{5} c_{5} \bigl(b_{4} - b_{5}\bigr)^{2}}{2}\,\dd(\sin\xi\sin\psi)\Big] \wedge\Big[ -\cos\tau\tan\xi\sin2\zeta\,\dd\xi\nonumber\\
&+(\cos\tau\cos2\zeta+\cos3\tau)\,\dd\zeta -\sin\tau\sin2\zeta\,\dd\tau\Big] +\cos\tau\Big[\bigl[ - X a_{1} a_{5} b_{5} c_{1} c_{4}\nonumber\\
&- Y a_{5}^{2} \bigl(b_5+b_1^2b_4\bigr)\bigr]\,\dd(\sin\xi\cos\psi) +Z c_{4} \bigl(b_{1}^{2} - b_{5}^{2}\bigr) \bigl(b_{4} - b_{5}\bigr)\,\dd(\sin\xi\sin\psi)\Big] \wedge\Big[\tan\xi\sin2\zeta\,\dd\xi\nonumber\\
&+(\cos2\tau-\cos2\zeta)\,\dd\zeta\Big] +\cos\tau\Big[Z a_{5} \bigl(b_{1}^{2} - b_{5}^{2}\bigr) \bigl(b_{4} - b_{5}\bigr)\,\dd(\sin\xi\cos\psi) +\bigl[X c_{4} c_{5} \bigl(b_4+b_1^2b_5\bigr)\nonumber\\
&+ Y a_{1} a_{5} b_{4} c_{1} c_{5}\bigr]\,\dd(\sin\xi\sin\psi)\Big] \wedge\Big[\sin2\tau\,\dd\zeta +\sin2\zeta\,\dd\tau\Big] +\frac{a_1\sin\xi}{c_1}\Big\{ -\sin\tau\Big[X a_{1} b_{4} c_{1} c_{5}^{2} \cos\psi\nonumber\\
&+ Y a_{4} c_{5} \cos\psi \bigl(b_4+b_1^2b_5\bigr) + Z a_{4} \sin\psi \bigl(b_{1}^{2} - b_{5}^{2}\bigr) \bigl(b_{4} - b_{5}\bigr)\Big] \Big[ -\tan\xi\sin2\tau\,\dd\xi \wedge\dd\zeta +\tan\xi\sin2\zeta\,\dd\xi\nonumber\\
&\wedge\dd\tau -(\cos2\zeta+\cos2\tau)\,\dd\zeta \wedge\dd\tau\Big] +\cos\tau\Big[ - X \sin\psi b_5\bigl[a_1c_1(1+b_4b_5) +\tfrac12(b_4-b_5)^2\bigr]\nonumber\\
&- Y a_{1} a_{4} a_{5} b_{5} c_{1} \sin\psi + \frac{Z b_{5} c_{5} \cos\psi \bigl(b_{4} - b_{5}\bigr)^{2}}{2}\Big] \Big[ -\tan\xi\sin2\tau\,\dd\xi \wedge\dd\zeta +\tan\xi\sin2\zeta\,\dd\xi \wedge\dd\tau\nonumber\\
&-(\cos2\zeta+\cos2\tau)\,\dd\zeta \wedge\dd\tau\Big] -\sin\tau\Big[ - X a_{5} c_{5} \sin\psi \bigl(b_5+b_1^2b_4\bigr) - Y a_{1} a_{4} a_{5} b_{5} c_{1} \sin\psi\nonumber\\
&+ Z c_{5} \cos\psi \bigl(b_{1}^{2} - b_{5}^{2}\bigr) \bigl(b_{4} - b_{5}\bigr)\Big] \Big[2\tan\xi\cos2\tau\,\dd\xi \wedge\dd\zeta -2\sin2\tau\,\dd\zeta \wedge\dd\tau\Big] +\Big[ - X a_{1} b_{5} c_{1} c_{4} c_{5} \cos\psi\nonumber\\
&- Y \cos\psi b_5\bigl[a_1c_1(1+b_4b_5) +\tfrac12(b_4-b_5)^2\bigr] + \frac{Z a_{5} b_{5} \sin\psi \bigl(b_{4} - b_{5}\bigr)^{2}}{2}\Big] \Big[\cos\tau\big[2\tan\xi\cos2\tau\,\dd\xi \wedge\dd\zeta\nonumber\\
&-2\sin2\tau\,\dd\zeta \wedge\dd\tau\big] -\sin\tau\big[\tan\xi\sin2\tau\,\dd\xi \wedge\dd\zeta +\tan\xi\sin2\zeta\,\dd\xi \wedge\dd\tau -(\cos2\zeta-\cos2\tau)\,\dd\zeta \wedge\dd\tau\big] \Big]\nonumber\\
&+\cos\tau\Big[X a_{5} c_{4} \sin\psi \bigl(b_4+b_1^2b_5\bigr) + Y a_{1} a_{5}^{2} b_{4} c_{1} \sin\psi + Z c_{4} \cos\psi \bigl(b_{1}^{2} - b_{5}^{2}\bigr) \bigl(b_{4} - b_{5}\bigr)\Big] \Big[\tan\xi\sin2\tau\,\dd\xi\nonumber\\
&\wedge\dd\zeta +\tan\xi\sin2\zeta\,\dd\xi \wedge\dd\tau -(\cos2\zeta-\cos2\tau)\,\dd\zeta \wedge\dd\tau\Big] \Big\}\Bigg\} +\frac{b_4-b_5}{2}\bigl[2b_1^2 -b_5(b_4+b_5)\bigr] \Bigg\{\nonumber\\
&-c_1\cos^3\xi\sin\zeta\cos^2\zeta\,\dd(\sin\xi\cos\psi) \wedge \Bigl[\sin\xi(a_5\sin\psi\cos\tau-c_5\cos\psi\sin\tau)\,\dd(\sin\xi\sin\psi)\nonumber\\
&+a_5\cos\xi\cos\zeta\,\dd(\cos\xi\cos\zeta\cos\tau)\Bigr] +\cos^3\xi\cos^2\zeta\,\dd(-\cos\xi\cos\zeta\sin\tau) \wedge \Bigl[\nonumber\\
&-c_1c_5\cos\xi\sin\zeta\cos\zeta\,\dd(\sin\xi\sin\psi) +a_1\sin\xi\cos\xi\sin\tau(a_5\sin\psi\sin\tau+c_5\cos\psi\cos\tau)\,\dd\zeta\nonumber\\
&+a_1\sin\xi\sin\zeta(a_5\sin\psi\cos\tau-c_5\cos\psi\sin\tau)\,\dd(\cos\xi\cos\zeta\cos\tau)\Bigr] \Bigg\}.
\label{eq:SO3FluxC22}
\end{align}
\endgroup
The right-hand side for
$C_{23}+\epsilon_{3jk}\mathcal A^j\wedge C_1{}^k
+\frac12C_0\epsilon_{3jk}\mathcal A^j\wedge\mathcal A^k$
is obtained from \eqref{eq:SO3FluxC22} by \eqref{eq:SO3FluxExchange}.

Finally, the three-form $C_3$ is determined by
\begingroup
\renewcommand{\baselinestretch}{1}
\small
\setlength{\jot}{2pt}
\thinmuskip=2mu
\medmuskip=3mu
\thickmuskip=4mu
\allowdisplaybreaks[4]
\begin{align}
&C_3 +\mathcal A^i \wedge C_{2i} +\frac12\epsilon_{ijk}\mathcal A^i \wedge\mathcal A^j \wedge C_1{}^k +\frac16C_0\epsilon_{ijk}\mathcal A^i \wedge\mathcal A^j \wedge\mathcal A^k ={}\frac{2\tan\xi}{\sin2\zeta}\,\dd\xi \wedge \left(\frac{c_1}{a_1}\dd\psi-\dd\tau\right)\nonumber\\
&\wedge(C_1{}^1+C_0\mathcal A^1) +\frac{b_1(b_4-b_5)^2\cos^3\xi}{4a_1}\Bigg\{ \sin\xi\sin2\zeta \big[(c_5X-a_5Y)\sin2\tau -2a_5c_5Z\cos2\tau\big] \dd\xi \wedge\dd\zeta\nonumber\\
&\wedge\dd\psi -\frac{a_1}{2}\sin\xi\cos^2\xi\sin2\zeta \big[(a_5^2+c_5^2)\sin2\psi\cos2\tau +(c_5^2-a_5^2)\sin2\psi\cos2\zeta -2a_5c_5\cos2\psi\sin2\tau\big] \dd\xi\nonumber\\
&\wedge\dd\zeta \wedge\dd\tau +\sin\xi\Big[(c_5X+a_5Y)(1+\cos^22\zeta) +2\cos2\zeta\big((c_5X-a_5Y)\cos2\tau+2a_5c_5Z\sin2\tau\big)\nonumber\\
&-c_1a_5c_5\cos^2\xi\sin^22\zeta\Big] \dd\xi \wedge\dd\psi \wedge\dd\tau +\cos\xi\sin2\zeta \big[(c_5X+a_5Y)\cos2\zeta +(c_5X-a_5Y)\cos2\tau\nonumber\\
&+2a_5c_5Z\sin2\tau\big] \dd\zeta \wedge\dd\psi \wedge\dd\tau\Bigg\}.
\label{eq:SO3FluxC3}
\end{align}
\endgroup

Evaluated at the SO(3)-invariant vacuum values (\ref{Field_Identifications}), the $D=11$ configuration (\ref{eq: upliftansatz}) with
(\ref{eq:SO3XYZ})--(\ref{eq:SO3FluxDecompositionMain}) and the quantities defined in sections~\ref{sec:MetCoefsSO3} and~\ref{sec:SO3InternalThreeForm} solves all $D=11$ field equations and thus provides a new $\cN=1$ SO(3)-invariant solution of $D=11$ supergravity.

\subsection{Recovering previously known supersymmetric solutions}
\label{sec:SO3RoundJoin}

We now recover the previously known supersymmetric $D=11$ solutions in the sector
(\ref{eq:SO3Identifications}), with enhanced SO$(8)$, G$_2$ and
$\mathrm{SU}(3)\times\mathrm{U}(1)$ symmetries.

At the SO(8)-invariant vacuum in \eqref{Field_Identifications},
$a_i=c_i=1$ and $b_i=0$ for $i=1,4,5$, and (\ref{eq: upliftansatz}) with
(\ref{eq:SO3XYZ})--(\ref{eq:SO3FluxDecompositionMain}) reduces to the Freund--Rubin solution. Indeed, the warp factor
\eqref{eq:SO3Bergerwarp} becomes $\Delta=1$, although the functions (\ref{eq:SO3XYZ}) remain non-trivial. The metric coefficients (\ref{eq:SO3Bergerorbitblocks})--(\ref{eq:SO3Bergerbase}) reduce to
\begin{gather}
h_{ab}=\operatorname{diag}\bigl(1,\cos^2\xi,\sin^2\xi,\cos^2\xi\cos^22\zeta\bigr)_{ab},\quad
 \gamma_{ij}=\cos^2\xi\,\operatorname{diag}(1,\cos^2\zeta,\sin^2\zeta)_{ij},\nonumber\\[4pt]
\mathcal A^{1}=-\sin2\zeta\,\dd\tau,\qquad
\mathcal A^{2}=\mathcal A^{3}=0, 
\label{eq:SO3BergerRoundBlocks}
\end{gather}
and the forms (\ref{eq:SO3FluxBaseComponents})--(\ref{eq:SO3FluxC3}) all vanish. Substitution of (\ref{eq:SO3BergerRoundBlocks}) into \eqref{eq:introBerger} gives the round seven-sphere
as the join of $S^1$ and $S^5$, equation (\ref{eq:SO3RoundJoin}) below. In order to see this, it is convenient to absorb the five angles
$(\zeta,\tau,\alpha,\beta,\gamma)$ into embedding coordinates
$\tilde\mu^i$, $i=1,\ldots,6$, on a unit $S^5$, with $\delta_{ij}\tilde\mu^i\tilde\mu^j=1$. The parametrisation
(\ref{eq:SO3Bergerslice}) then reads
\begin{equation}\label{eq:SO3RoundEmbedding}
\begin{aligned}
-\mu^1+i\mu^8&=\sin\xi\,e^{i\psi},\\
\mu^3+i\mu^2&=\cos\xi
\left(\tilde\mu^1+i\tilde\mu^2\right),\\
\mu^5+i\mu^4&=\cos\xi
\left(\tilde\mu^3+i\tilde\mu^4\right),\\
\mu^7+i\mu^6&=\cos\xi
\left(\tilde\mu^5+i\tilde\mu^6\right) \; .
\end{aligned}
\end{equation}
Then, (\ref{eq:introBerger}) with (\ref{eq:SO3BergerRoundBlocks}), (\ref{eq:SO3RoundEmbedding}) indeed gives
\begin{equation}\label{eq:SO3RoundJoin}
\dd s_7^2=g^{-2}\left[
\dd\xi^2+\sin^2\xi\,\dd\psi^2
+\cos^2\xi\,\delta_{ij}\,\dd\tilde\mu^i\dd\tilde\mu^j
\right].
\end{equation}

To recover the $\cN=1$ G$_2$-invariant solution, it is convenient
to replace (\ref{eq:SO3RoundEmbedding}) with
\begin{equation}\label{eq:SO3SphereEmbedding}
\begin{aligned}
\mu^1&=-\sin\vartheta\cos\eta,\\
\mu^3+i\mu^2&=\sin\vartheta\sin\eta
\left(\tilde\mu^1+i\tilde\mu^2\right),\\
\mu^5+i\mu^4&=\sin\vartheta\sin\eta
\left(\tilde\mu^3+i\tilde\mu^4\right),\\
\mu^7+i\mu^6&=\sin\vartheta\sin\eta
\left(\tilde\mu^5+i\tilde\mu^6\right),\\
\mu^8&=\cos\vartheta,
\end{aligned}
\end{equation}
with $\delta_{ij}\tilde\mu^i\tilde\mu^j=1$ as before.
Evaluating (\ref{eq:SO3Bergerwarp})--(\ref{eq:SO3FluxDecompositionMain})
at the G$_2$-invariant vacuum in (\ref{Field_Identifications}),
and using the identities in appendix B.3 of \cite{Larios:2019kbw},
we find
\begin{equation}\label{eq:SO3G2Fields}
\begin{aligned}
\Delta^{-3}={}&
\frac{216}{125}\sqrt{\frac{2\sqrt3}{5}}\,
(2+\cos2\vartheta)^2,\\[3pt]
\dd s_7^2={}&
\frac1{g^2}\left(\frac{25}{12}\right)^{1/6}
(2+\cos2\vartheta)^{2/3}
\left[
\frac13\dd\vartheta^2
+\frac{\sin^2\vartheta}{2+\cos2\vartheta}\dd s^2(S^6)
\right],\\[3pt]
A_3={}&-\frac{\sin^2\vartheta}
{\sqrt2\,3^{3/4}g^3(2+\cos2\vartheta)}
\Big[
(2+\cos2\vartheta)\mathcal{J}\wedge\dd\vartheta\\
&\hspace{22mm}
+3\sin\vartheta\cos\vartheta\,\operatorname{Re}\Upomega
+\sqrt3\sin^2\vartheta\,\operatorname{Im}\Upomega
\Big].
\end{aligned}
\end{equation}
Here
$\dd s^2(S^6)=\dd\eta^2+\sin^2\eta\,
\delta_{ij}\dd\tilde\mu^i\dd\tilde\mu^j$
is the unit round six-sphere metric, and
$(\mathcal{J},\Upomega)$ is its homogeneous nearly-K\"ahler
structure, satisfying
$\dd\mathcal{J}=3\operatorname{Re}\Upomega$ and
$\dd\operatorname{Im}\Upomega=-2\mathcal{J}\wedge\mathcal{J}$.
For comparison with \cite{Larios:2019kbw}, we identify
\begin{equation}\label{eq:SO3G2Conventions}
\begin{gathered}
\beta_{\textrm{there}}=\vartheta_{\textrm{here}},\\
(\mu^1,\ldots,\mu^8)_{\textrm{there}}
=(\mu^2,-\mu^3,\mu^5,\mu^4,\mu^7,\mu^6,-\mu^1,\mu^8)_{\textrm{here}}.
\end{gathered}
\end{equation}
The resulting eleven-dimensional solution (\ref{eq: upliftansatz}) with (\ref{eq:SO3G2Fields}) reproduces (4.2), (4.3)
of \cite{Larios:2019kbw}, with the opposite pseudoscalar branch:
$\chi_{\textrm{here}}=-\chi_{\textrm{there}}=-12^{-1/4}$.
Accordingly, the metrics and external four-forms agree, while
$A_{3,\textrm{here}}=-A_{\textrm{there}}$,
where $A_{\textrm{there}}$ is the potential in its equation (4.3).

Finally, the $\cN=2$ CPW solution \cite{Corrado:2001nv} is recovered using the
parametrisation (\ref{eq:SO3RoundEmbedding}). Let
$(\bm{\eta}^{(5)},\bm{J}^{(5)},\bm{\Omega}^{(5)})$ be the
Sasaki--Einstein structure constructed from $\tilde\mu^i$
as in appendix B.1 of \cite{Larios:2019kbw}, with
$\delta_{ij}\dd\tilde\mu^i\dd\tilde\mu^j = ds^2(\mathbb{CP}^2) + (\bm{\eta}^{(5)})^2 $. Defining
\begin{equation}\label{eq:SO3CPWForms}
\bm{\eta}'=\bm{\eta}^{(5)}+\frac13\dd\psi,
\qquad
\bm{J}'=\bm{J}^{(5)},
\qquad
\bm{\Omega}'=-i e^{i\psi}\bm{\Omega}^{(5)},
\end{equation}
the expressions above evaluated at the $\textrm{SU}(3) \times \textrm{U}(1)$ vacuum in
(\ref{Field_Identifications}) give
\begin{equation}\label{eq:SO3CPWFields}
\begin{aligned}
\Delta^{-3}={}&
\frac{3\sqrt3}{4}(1+2\sin^2\xi)^2,\\[3pt]
\dd s_7^2={}&
\frac{(1+2\sin^2\xi)^{2/3}}{3\,2^{2/3}g^2}
\Bigg[
2\dd\xi^2
+\frac{6\cos^2\xi}{1+2\sin^2\xi}\dd s^2(\mathbb{CP}^2)
+\frac{18\sin^2\xi\cos^2\xi}{1+8\sin^4\xi}
(\bm{\eta}')^2\\
&\hspace{15mm}
+\frac{1+8\sin^4\xi}{(1+2\sin^2\xi)^2}
\left(\dd\psi-\frac{3\cos^2\xi}{1+8\sin^4\xi}
\bm{\eta}'\right)^2
\Bigg],\\[3pt]
A_3={}&
\frac{\cos^2\xi}{\sqrt3\,g^3}
\left[
\dd\xi\wedge\operatorname{Im}\bm{\Omega}'
+\frac{\sin\xi\cos\xi}{1+2\sin^2\xi}
(2\dd\psi-3\bm{\eta}')
\wedge\operatorname{Re}\bm{\Omega}'
\right].
\end{aligned}
\end{equation}
These reproduce the CPW solution \cite{Corrado:2001nv} as given in (3.27)--(3.28)
of \cite{Larios:2019kbw}, with the four-dimensional graviphoton
switched off, under the identifications
\begin{equation}\label{eq:SO3CPWComparison}
\begin{aligned}
\alpha_{\textrm{there}}&=\xi_{\textrm{here}},
\qquad
\psi'_{\textrm{there}}=\psi_{\textrm{here}},\\
\left(\tilde\mu^{2a-1}+i\tilde\mu^{2a}\right)_{\textrm{there}}
&=e^{-i\pi/4}
\left(\tilde\mu^{2a-1}+i\tilde\mu^{2a}\right)_{\textrm{here}},
\qquad a=1,2,3.
\end{aligned}
\end{equation}
The constant rotation leaves $\bm{\eta}^{(5)}$, $\bm{J}^{(5)}$
and the metric unchanged, while
$\bm{\Omega}^{(5)}_{\textrm{there}}
=e^{-3i\pi/4}\bm{\Omega}^{(5)}_{\textrm{here}}$.
Thus (\ref{eq:SO3CPWForms}) agrees with the phase convention
$\bm{\Omega}'=e^{i(\psi+\pi/4)}\bm{\Omega}^{(5)}_{\textrm{there}}$
of that reference, and the internal three-forms coincide
without a gauge transformation. The external metric normalisation in that reference is related to
our unit-radius AdS$_4$ metric by
$\dd\bar s^2_{4,\textrm{there}}=g^{-2}\dd s^2(\mathrm{AdS}_4)$,
as follows from (2.46) of \cite{Larios:2019kbw}. 

\newpage

\bibliography{references}

\providecommand{\href}[2]{#2}\begingroup\raggedright\begin{thebibliography}{10}

\bibitem{deWit:1981sst}
B.~de~Wit and H.~Nicolai, {\it {N=8 Supergravity with Local SO(8) x SU(8)
  Invariance}},  {\em Phys. Lett. B} {\bf 108} (1982) 285.

\bibitem{deWit:1982ig}
B.~de~Wit and H.~Nicolai, {\it {N=8 Supergravity}},  {\em Nucl.Phys.} {\bf
  B208} (1982) 323.

\bibitem{Cremmer:1978km}
E.~Cremmer, B.~Julia, and J.~Scherk, {\it {Supergravity Theory in
  Eleven-Dimensions}},  {\em Phys.Lett.} {\bf B76} (1978) 409--412.

\bibitem{deWit:1983vq}
B.~de~Wit and H.~Nicolai, {\it {On the Relation Between $d=4$ and $d=11$
  Supergravity}},  {\em Nucl. Phys. B} {\bf 243} (1984) 91--111.

\bibitem{deWit:1984nz}
B.~de~Wit, H.~Nicolai, and N.~P. Warner, {\it {The Embedding of Gauged $N=8$
  Supergravity Into $d=11$ Supergravity}},  {\em Nucl. Phys.} {\bf B255} (1985)
  29.

\bibitem{deWit:1986iy}
B.~de~Wit and H.~Nicolai, {\it {The Consistency of the $S^7$ Truncation in
  $D=11$ Supergravity}},  {\em Nucl.Phys.} {\bf B281} (1987) 211.

\bibitem{Duff:1986hr}
M.~J. Duff, B.~E.~W. Nilsson, and C.~N. Pope, {\it {Kaluza-Klein
  Supergravity}},  {\em Phys. Rept.} {\bf 130} (1986) 1--142.

\bibitem{Nicolai:2011cy}
H.~Nicolai and K.~Pilch, {\it {Consistent Truncation of d = 11 Supergravity on
  AdS$_4 \times S^7$}},  {\em JHEP} {\bf 1203} (2012) 099,
  [\href{http://arxiv.org/abs/1112.6131}{{\tt arXiv:1112.6131}}].

\bibitem{deWit:2013ija}
B.~de~Wit and H.~Nicolai, {\it {Deformations of gauged SO(8) supergravity and
  supergravity in eleven dimensions}},  {\em JHEP} {\bf 05} (2013) 077,
  [\href{http://arxiv.org/abs/1302.6219}{{\tt arXiv:1302.6219}}].

\bibitem{Godazgar:2013nma}
H.~Godazgar, M.~Godazgar, and H.~Nicolai, {\it {Testing the non-linear flux
  ansatz for maximal supergravity}},  {\em Phys. Rev. D} {\bf 87} (2013)
  085038, [\href{http://arxiv.org/abs/1303.1013}{{\tt arXiv:1303.1013}}].

\bibitem{Lee:2014mla}
K.~Lee, C.~Strickland-Constable, and D.~Waldram, {\it {Spheres, generalised
  parallelisability and consistent truncations}},  {\em Fortsch. Phys.} {\bf
  65} (2017), no.~10-11 1700048, [\href{http://arxiv.org/abs/1401.3360}{{\tt
  arXiv:1401.3360}}].

\bibitem{Varela:2015ywx}
O.~Varela, {\it {Complete $D=11$ embedding of SO(8) supergravity}},  {\em Phys.
  Rev.} {\bf D97} (2018), no.~4 045010,
  [\href{http://arxiv.org/abs/1512.04943}{{\tt arXiv:1512.04943}}].

\bibitem{Kruger:2016agp}
O.~Kr{\"u}ger, {\it {Non-linear uplift Ans{\"a}tze for the internal metric and
  the four-form field-strength of maximal supergravity}},  {\em JHEP} {\bf 05}
  (2016) 145, [\href{http://arxiv.org/abs/1602.03327}{{\tt arXiv:1602.03327}}].

\bibitem{Comsa:2019rcz}
I.~M. Comsa, M.~Firsching, and T.~Fischbacher, {\it {SO(8) Supergravity and the
  Magic of Machine Learning}},  {\em JHEP} {\bf 08} (2019) 057,
  [\href{http://arxiv.org/abs/1906.00207}{{\tt arXiv:1906.00207}}].

\bibitem{Warner:1983vz}
N.~Warner, {\it {Some New Extrema of the Scalar Potential of Gauged $N=8$
  Supergravity}},  {\em Phys.Lett.} {\bf B128} (1983) 169.

\bibitem{Fischbacher:2009cj}
T.~Fischbacher, {\it {Fourteen new stationary points in the scalar potential of
  SO(8)-gauged N=8, D=4 supergravity}},  {\em JHEP} {\bf 09} (2010) 068,
  [\href{http://arxiv.org/abs/0912.1636}{{\tt arXiv:0912.1636}}].

\bibitem{Fischbacher:2010ec}
T.~Fischbacher, K.~Pilch, and N.~P. Warner, {\it {New Supersymmetric and
  Stable, Non-Supersymmetric Phases in Supergravity and Holographic Field
  Theory}},  \href{http://arxiv.org/abs/1010.4910}{{\tt arXiv:1010.4910}}.

\bibitem{Bobev:2019dik}
N.~Bobev, T.~Fischbacher, and K.~Pilch, {\it {Properties of the new $
  \mathcal{N} $ = 1 AdS$_{4}$ vacuum of maximal supergravity}},  {\em JHEP}
  {\bf 01} (2020) 099, [\href{http://arxiv.org/abs/1909.10969}{{\tt
  arXiv:1909.10969}}].

\bibitem{Freund:1980xh}
P.~G. Freund and M.~A. Rubin, {\it {Dynamics of Dimensional Reduction}},  {\em
  Phys.Lett.} {\bf B97} (1980) 233--235.

\bibitem{Corrado:2001nv}
R.~Corrado, K.~Pilch, and N.~P. Warner, {\it {An N=2 supersymmetric membrane
  flow}},  {\em Nucl. Phys.} {\bf B629} (2002) 74--96,
  [\href{http://arxiv.org/abs/hep-th/0107220}{{\tt hep-th/0107220}}].

\bibitem{Cvetic:2000yp}
M.~Cvetic, H.~Lu, and C.~N. Pope, {\it {Consistent warped space Kaluza-Klein
  reductions, half maximal gauged supergravities and $\mathbb{CP}^{n}$
  constructions}},  {\em Nucl. Phys.} {\bf B597} (2001) 172--196,
  [\href{http://arxiv.org/abs/hep-th/0007109}{{\tt hep-th/0007109}}].

\bibitem{Larios:2019kbw}
G.~Larios, P.~Ntokos, and O.~Varela, {\it {Embedding the SU(3) sector of SO(8)
  supergravity in $D=11$}},  {\em Phys.\ Rev.\ D} {\bf 100} (2019), no.~8
  086021, [\href{http://arxiv.org/abs/1907.02087}{{\tt arXiv:1907.02087}}].

\bibitem{Hull:1984yy}
C.~Hull, {\it {New Gauging of $N=8$ Supergravity}},  {\em Phys.Rev.} {\bf D30}
  (1984) 760.

\bibitem{DallAgata:2011aa}
G.~Dall'Agata and G.~Inverso, {\it {On the Vacua of N = 8 Gauged Supergravity
  in 4 Dimensions}},  {\em Nucl.Phys.} {\bf B859} (2012) 70--95,
  [\href{http://arxiv.org/abs/1112.3345}{{\tt arXiv:1112.3345}}].

\bibitem{Guarino:2015qaa}
A.~Guarino and O.~Varela, {\it {Dyonic ISO(7) supergravity and the duality
  hierarchy}},  {\em JHEP} {\bf 02} (2016) 079,
  [\href{http://arxiv.org/abs/1508.04432}{{\tt arXiv:1508.04432}}].

\bibitem{Guarino:2019snw}
A.~Guarino, J.~Tarrio, and O.~Varela, {\it {Flowing to $\mathcal{N}=3$
  Chern-Simons-matter theory}},  {\em JHEP} {\bf 03} (2020) 100,
  [\href{http://arxiv.org/abs/1910.06866}{{\tt arXiv:1910.06866}}].

\bibitem{Duff:2006ue}
M.~J. Duff and S.~Ferrara, {\it {E(7) and the tripartite entanglement of seven
  qubits}},  {\em Phys. Rev. D} {\bf 76} (2007) 025018,
  [\href{http://arxiv.org/abs/quant-ph/0609227}{{\tt quant-ph/0609227}}].

\bibitem{Levay:2006pt}
P.~Levay, {\it {Strings, black holes, the tripartite entanglement of seven
  qubits and the Fano plane}},  {\em Phys. Rev. D} {\bf 75} (2007) 024024,
  [\href{http://arxiv.org/abs/hep-th/0610314}{{\tt hep-th/0610314}}].

\bibitem{Duff:2010vy}
M.~J. Duff and S.~Ferrara, {\it {Four curious supergravities}},  {\em Phys.
  Rev. D} {\bf 83} (2011) 046007, [\href{http://arxiv.org/abs/1010.3173}{{\tt
  arXiv:1010.3173}}].

\bibitem{Ferrara:2016fwe}
S.~Ferrara and R.~Kallosh, {\it {Seven-disk manifold, $\alpha$-attractors, and
  $B$ modes}},  {\em Phys. Rev. D} {\bf 94} (2016), no.~12 126015,
  [\href{http://arxiv.org/abs/1610.04163}{{\tt arXiv:1610.04163}}].

\bibitem{Gunaydin:2020ric}
M.~Gunaydin, R.~Kallosh, A.~Linde, and Y.~Yamada, {\it {M-theory Cosmology,
  Octonions, Error Correcting Codes}},  {\em JHEP} {\bf 01} (2021) 160,
  [\href{http://arxiv.org/abs/2008.01494}{{\tt arXiv:2008.01494}}].

\bibitem{Baez:2001dm}
J.~C. Baez, {\it {The Octonions}},  {\em Bull. Am. Math. Soc.} {\bf 39} (2002)
  145--205, [\href{http://arxiv.org/abs/math/0105155}{{\tt math/0105155}}].
  [Erratum: Bull.Am.Math.Soc. 42, 213 (2005)].

\bibitem{Cvetic:2015uga}
M.~Cvetic, G.~Gibbons, and C.~Pope, {\it {Compactifications of Deformed
  Conifolds, Branes and the Geometry of Qubits}},  {\em JHEP} {\bf 01} (2016)
  135, [\href{http://arxiv.org/abs/1507.07585}{{\tt arXiv:1507.07585}}].

\bibitem{Awada:1983Squashed}
M.~A. Awada, M.~J. Duff, and C.~N. Pope, {\it {$N=8$ Supergravity Breaks Down
  to $N=1$}},  {\em Phys. Rev. Lett.} {\bf 50} (1983) 294--297.

\bibitem{Englert:1983rn}
F.~Englert and H.~Nicolai, {\it {Supergravity in eleven-dimensional
  space-time}},  in {\em {12th International Colloquium on Group Theoretical
  Methods in Physics}} (G.~Denardo, G.~Ghirardi, and T.~Weber, eds.), vol.~201
  of {\em Lecture Notes in Physics}, (Berlin), pp.~249--283, Springer, 1984.
\newblock Trieste, Italy, September 1983.

\bibitem{Sezgin:1983ik}
E.~Sezgin, {\it {The Spectrum of the Eleven-dimensional Supergravity
  Compactified on the Round Seven Sphere}},  {\em Phys. Lett. B} {\bf 138}
  (1984) 57--62.

\bibitem{Biran:1983iy}
B.~Biran, A.~Casher, F.~Englert, M.~Rooman, and P.~Spindel, {\it {The
  Fluctuating Seven Sphere in Eleven-dimensional Supergravity}},  {\em Phys.
  Lett. B} {\bf 134} (1984) 179--183.

\bibitem{Nicolai:1985hs}
H.~Nicolai and N.~P. Warner, {\it {The SU(3) X U(1) Invariant Breaking of
  Gauged $N=8$ Supergravity}},  {\em Nucl. Phys.} {\bf B259} (1985) 412.

\bibitem{Klebanov:2008vq}
I.~Klebanov, T.~Klose, and A.~Murugan, {\it {AdS(4)/CFT(3) Squashed, Stretched
  and Warped}},  {\em JHEP} {\bf 03} (2009) 140,
  [\href{http://arxiv.org/abs/0809.3773}{{\tt arXiv:0809.3773}}].

\bibitem{Malek:2019eaz}
E.~Malek and H.~Samtleben, {\it {Kaluza-Klein Spectrometry for Supergravity}},
  {\em Phys. Rev. Lett.} {\bf 124} (2020), no.~10 101601,
  [\href{http://arxiv.org/abs/1911.12640}{{\tt arXiv:1911.12640}}].

\bibitem{Malek:2020yue}
E.~Malek and H.~Samtleben, {\it {Kaluza-Klein Spectrometry from Exceptional
  Field Theory}},  {\em Phys. Rev. D} {\bf 102} (2020), no.~10 106016,
  [\href{http://arxiv.org/abs/2009.03347}{{\tt arXiv:2009.03347}}].

\bibitem{Varela:2020Spectra}
O.~Varela, {\it {Super-Chern-Simons spectra from Exceptional Field Theory}},
  {\em JHEP} {\bf 04} (2021) 283, [\href{http://arxiv.org/abs/2010.09743}{{\tt
  arXiv:2010.09743}}].

\bibitem{Cesaro:2020soq}
M.~Ces{\`a}ro and O.~Varela, {\it {Kaluza-Klein fermion mass matrices from
  Exceptional Field Theory and ${\cal N}=1$ spectra}},  {\em JHEP} {\bf 03}
  (2021) 138, [\href{http://arxiv.org/abs/2012.05249}{{\tt arXiv:2012.05249}}].

\bibitem{Cesaro:2021KK}
M.~Ces{\`a}ro, G.~Larios, and O.~Varela, {\it {Supersymmetric spectroscopy on
  $\mathrm{AdS}_4\times S^7$ and $\mathrm{AdS}_4\times S^6$}},  {\em JHEP} {\bf
  07} (2021) 094, [\href{http://arxiv.org/abs/2103.13408}{{\tt
  arXiv:2103.13408}}]. Erratum: JHEP 05 (2024) 084.

\bibitem{BenettiGenolini:2023kxp}
P.~Benetti~Genolini, J.~P. Gauntlett, and J.~Sparks, {\it {Equivariant
  Localization in Supergravity}},  {\em Phys. Rev. Lett.} {\bf 131} (2023),
  no.~12 121602, [\href{http://arxiv.org/abs/2306.03868}{{\tt
  arXiv:2306.03868}}].

\bibitem{BenettiGenolini:2023AdSCFT}
P.~Benetti~Genolini, J.~P. Gauntlett, and J.~Sparks, {\it {Equivariant
  localization for AdS/CFT}},  {\em JHEP} {\bf 02} (2024) 015,
  [\href{http://arxiv.org/abs/2308.11701}{{\tt arXiv:2308.11701}}].

\bibitem{Couzens:2026TenDimensional}
C.~Couzens, A.~L{\"u}scher, and J.~Sparks, {\it {Ten-dimensional
  localization}},  {\em Phys. Rev. D} {\bf 114} (2026), no.~8 086022,
  [\href{http://arxiv.org/abs/2607.05529}{{\tt arXiv:2607.05529}}].

\bibitem{Giombi:2023M2}
S.~Giombi and A.~A. Tseytlin, {\it {Wilson Loops at Large $N$ and the Quantum
  M2-Brane}},  {\em Phys. Rev. Lett.} {\bf 130} (2023) 201601,
  [\href{http://arxiv.org/abs/2303.15207}{{\tt arXiv:2303.15207}}].

\bibitem{Gautason:2025M2}
F.~F. Gautason and J.~van Muiden, {\it {Localization of the M2-Brane}},  {\em
  Phys. Rev. Lett.} {\bf 135} (2025) 101601,
  [\href{http://arxiv.org/abs/2503.16597}{{\tt arXiv:2503.16597}}].

\bibitem{Bena:2020BraneJets}
I.~Bena, K.~Pilch, and N.~P. Warner, {\it {Brane-Jet Instabilities}},  {\em
  JHEP} {\bf 10} (2020) 091, [\href{http://arxiv.org/abs/2003.02851}{{\tt
  arXiv:2003.02851}}].

\bibitem{Guarino:2020BraneJets}
A.~Guarino, J.~Tarr{\'i}o, and O.~Varela, {\it {Brane-jet stability of
  non-supersymmetric AdS vacua}},  {\em JHEP} {\bf 09} (2020) 110,
  [\href{http://arxiv.org/abs/2005.07072}{{\tt arXiv:2005.07072}}].

\bibitem{Guarino:2015jca}
A.~Guarino, D.~L. Jafferis, and O.~Varela, {\it {The string origin of dyonic
  N=8 supergravity and its simple Chern-Simons duals}},  {\em Phys. Rev. Lett.}
  {\bf 115} (2015), no.~9 091601, [\href{http://arxiv.org/abs/1504.08009}{{\tt
  arXiv:1504.08009}}].

\bibitem{Guarino:2015vca}
A.~Guarino and O.~Varela, {\it {Consistent $ \mathcal{N}=8 $ truncation of
  massive IIA on S$^{6}$}},  {\em JHEP} {\bf 12} (2015) 020,
  [\href{http://arxiv.org/abs/1509.02526}{{\tt arXiv:1509.02526}}].

\bibitem{Varela:2015uca}
O.~Varela, {\it {AdS$_{4}$ solutions of massive IIA from dyonic ISO(7)
  supergravity}},  {\em JHEP} {\bf 03} (2016) 071,
  [\href{http://arxiv.org/abs/1509.07117}{{\tt arXiv:1509.07117}}].

\bibitem{Bobev:2020qev}
N.~Bobev, T.~Fischbacher, F.~F. Gautason, and K.~Pilch, {\it {New AdS$_4$ Vacua
  in Dyonic ISO(7) Gauged Supergravity}},  {\em JHEP} {\bf 02} (2021) 215,
  [\href{http://arxiv.org/abs/2011.08542}{{\tt arXiv:2011.08542}}].

\bibitem{deWit:2007kvg}
B.~de~Wit, H.~Samtleben, and M.~Trigiante, {\it {The Maximal D=4
  supergravities}},  {\em JHEP} {\bf 06} (2007) 049,
  [\href{http://arxiv.org/abs/0705.2101}{{\tt arXiv:0705.2101}}].

\end{thebibliography}\endgroup
\end{document}